\documentclass[epjc,aps,showpacs,preprintnumbers,amsmath,nofootinbib,amssymb]{revtex4}
\usepackage[dvips]{epsfig,rotating,subeqnarray}

\newcommand{\nc}[1]{\newcommand{#1}}

\nc{\its}[1]{\itshape #1 \upshape}
\nc{\mc}[3]{\multicolumn{#1}{#2}{#3}}

\nc{\bc}{\begin{center}}
\nc{\ec}{\end{center}}
\nc{\ig}[1]{\bc \includegraphics{#1} \ec}

\nc{\bo}[1]{\mbox{\boldmath \( #1 \! \! \)  \unboldmath}}
\nc{\be}{\begin{eqnarray}}
\nc{\ee}{\end{eqnarray}}
\nc{\bew}{\begin{eqnarray*}}
\nc{\eew}{\end{eqnarray*}}
\nc{\bs}{\begin{subeqnarray}}   
\nc{\es}{\end{subeqnarray}}     
\nc{\nnn}{\nonumber \\}
\nc{\f}[2]{\frac{#1}{#2}}
\nc{\td}[2]{\f{d #1}{d #2}}
\nc{\pd}[2]{\f{\partial #1}{\partial #2}}
\nc{\suli}{\sum\limits}
\nc{\proli}{\prod\limits}
\nc{\ili}{\int\limits}
\nc{\sr}[2]{\stackrel{#1}{#2}}
\nc{\dps}{\displaystyle}
\nc{\ket}[1]{\left| #1 \right>}
\nc{\bra}[1]{\left< #1 \right|}
\nc{\bracket}[2]{\left< #1 \right| \left. \! #2 \right>}
\nc{\norm}[1]{\left\| #1 \right\|}
\nc{\lndm}[1]{\pd{^{#1} \ln{\det{M}}}{\mu^{#1}}}
\nc{\pdmm}[1]{M^{-1} \pd{^{#1} M}{\mu^{#1}}}
\nc{\pdm}{M^{-1}\pd{M}{\mu}}
\nc{\trac}[1]{\mbox{Tr}\left(#1\right)}

\def\lsim{\raise0.3ex\hbox{$<$\kern-0.75em\raise-1.1ex\hbox{$\sim$}}}
\def\gsim{\raise0.3ex\hbox{$>$\kern-0.75em\raise-1.1ex\hbox{$\sim$}}}

\begin{document}

\title{Screening of heavy quark free energies at finite temperature and \\
non-zero baryon chemical potential}

\author{M. D\"oring$^{\rm a, b}$, S. Ejiri$^{\rm c}$, O. Kaczmarek$^{\rm a}$,
  F. Karsch$^{\rm a, b}$, E. Laermann$^{\rm a}$}

\address{
$^{\rm a}$Fakult\"at f\"ur Physik, Universit\"at Bielefeld, D-33615 Bielefeld,
Germany\\
$^{\rm b}$Physics Department, Brookhaven National Laboratory, 
Upton, NYb11973, USA\\
$^{\rm c}$Department of Physics, The University of Tokyo,
\small Tokyo 113-0033, Japan}

\date{\today}

\preprint{BI-TP-2005-28 and BNL-NT-05-27 and TKYNT-05-17}

\begin{abstract}
We analyze the dependence of heavy quark free energies on
the baryon chemical potential $\mu_b$ in 2-flavour QCD using improved (p4)
staggered fermions with a bare quark mass of $\hat{m}/T = 0.4$. By performing 
a 6$^{th}$ order Taylor expansion in the chemical potential which 
circumvents the sign problem. The Taylor expansion coefficients 
of colour singlet and colour averaged free energies are calculated
and from this the expansion coefficients for the corresponding 
screening masses are determined. We find that for small $\mu_b$ the free
energies of a static quark anti-quark pair decrease in a medium with a net 
excess of quarks and that screening is well described by a screening 
mass which increases with increasing $\mu_b$. The 
$\mu_b$-dependent corrections to the screening masses are well described by 
perturbation theory for $T\gsim 2 T_c$. In particular, we find for all
temperatures above $T_c$ that the expansion coefficients for singlet and 
colour averaged screening masses differ by a factor 2.
\end{abstract}


\maketitle

\section{Introduction}
\label{intro}
Numerical studies of QCD provided quite detailed 
information about the properties of matter at high temperature and
vanishing net baryon density \cite{review}. In particular, the screening
of static quark anti-quark sources at large distances and their 
renormalization has been analyzed in quite some detail 
\cite{fthq,fthq2,pisarski}. Compared to this our knowledge on the
dependence of the equation of state and screening at non-zero baryon
number density, or equivalently, non-zero baryon chemical 
potential ($\mu_b$) is rather limited. The $\mu_b$-dependence of the 
QCD partition function \cite{vuorinen} and the HTL-resummation of the pressure
\cite{rebhan} have been evaluated only recently.
Although in leading order of high temperature perturbation theory the 
dependence of the Debye screening mass on 
$\mu_b$ is well-known \cite{debye-mu} neither the temperature range for the 
validity of this perturbative result nor the generic features of screening 
of heavy quark free energies at non-zero $\mu_b$ have been analyzed so far
with non-perturbative methods in the vicinity of the QCD transition \footnote{A
first attempt to calculate heavy quark free energies at non-zero quark
chemical potential has been discussed in \cite{fodor-debye}.}.

Recently studies of the equation of state have successfully been
extended to non-vanishing baryon chemical potential using Taylor 
expansions \cite{eos} around $\mu_b=0$ as well as reweighting 
techniques \cite{Fodor} and imaginary chemical potentials \cite{lombardo}. 
We will use here the former approach to analyze
the screening of static quark anti-quark sources at non-zero $\mu_b$,
{\it i.e.} in a medium with a non-vanishing net quark density.
We evaluate the Taylor expansion coefficients for correlation functions 
of heavy quark anti-quark pairs and deduce from this expansion 
coefficients for the screening mass. 

For this work we analyzed gauge field configurations using a p4-improved staggered fermion action with $N_f = 2$
degenerate quark flavours. We used the same data sample that recently
has been generated by the Bielefeld-Swansea collaboration for
the analysis of the equation of state \cite{eos}. This sample consists
of 1000 up to 4000 gauge field configurations available for several
bare gauge couplings below and above the transition temperature. The lattice
size is $16^3 \times 4$  and the bare quark mass, $\hat{m}/T=0.4$, corresponds to
a pion mass of about $770$~MeV. In addition we generated 1000 configurations 
at $T=3T_c$ and 1600 at $T=4T_c$ to check for the approach to the high
temperature perturbative regime. 
In addition to gauge invariant colour averaged free energies we also 
have analyzed colour singlet free energies. To do so all gauge configurations 
have been transformed to Coulomb gauge before evaluating the Polyakov loop 
correlation functions.

This paper is organized as follows. In the next section we discuss 
the general setup for calculating Taylor expansions for heavy quark
free energies. Our numerical results for singlet and colour averaged free 
energies are presented in section III. In section IV we discuss the
determination of $\mu_b$-dependent corrections to screening masses
from the free energies. Our conclusions are given in section V. In an
appendix we give detailed expressions for the Taylor expansion coefficients
of purely gluonic observables. 

\section{Taylor expansion of heavy quark free energies}
\label{taylor}

A heavy (static) quark $Q$ at site $\bo{x}$ is represented by the Polyakov 
loop, 
\be
L(\bo{x}) = \proli_{x_4 = 1}^{N_\tau} U_4(\bo{x}, x_4) \; , 
\ee
which is an SU(3) matrix.
A heavy anti-quark $\bar{Q}$ is described by the corresponding hermitian
conjugate matrix. The free energy of a $Q\bar{Q}$-pair separated by a 
distance $r$ is then
calculated from the expectation value of the correlation function of 
$L(\bo{0})$ and $L^\dagger(\bo{r})$ where $\bo{r}$ points to a site with
distance $r$ to $\bo{0}$. The dependence on the baryon chemical potential
$\mu_b$ or quark chemical potential $\mu\equiv   \mu_b/3$ is established solely
through the fermion determinant, $\det{M(\{ U_\rho(x) \}, \mu, \hat{m})}$, with
$\hat{m}$ denoting the bare quark mass. 
In order to avoid the sign problem, which arises from $Re\left(\det{M}
\right)$ not being positive definite for $\mu \ne 0$, 
Taylor expansions in the quark chemical potential 
are used. This allows us to perform our simulations at zero chemical potential
thereby restricting us to small chemical potentials.

\begin{figure}[t]
\bew
\begin{array}{cc}
\begin{turn}{270} \epsfbox{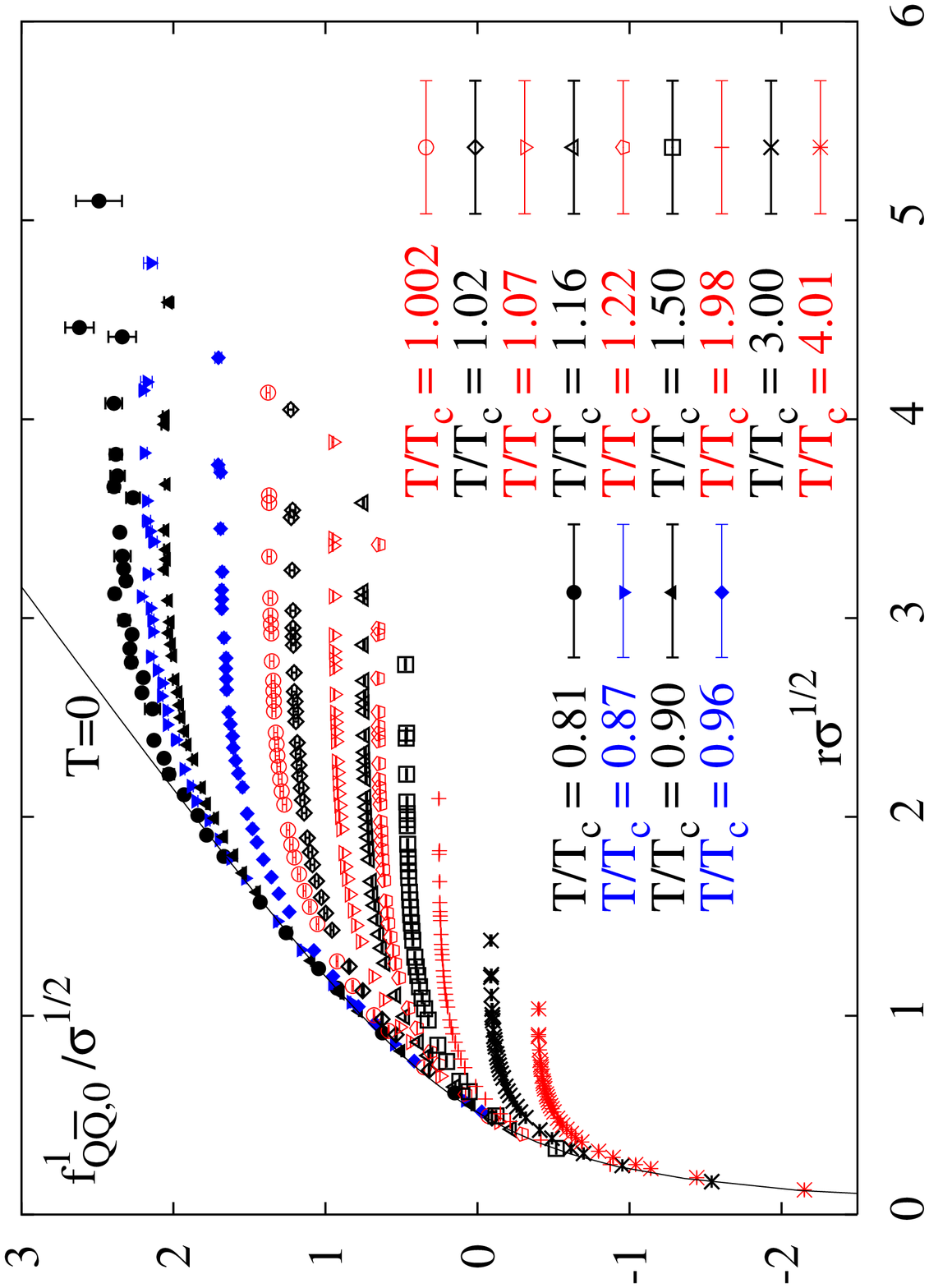} \end{turn} &
\begin{turn}{270} \epsfbox{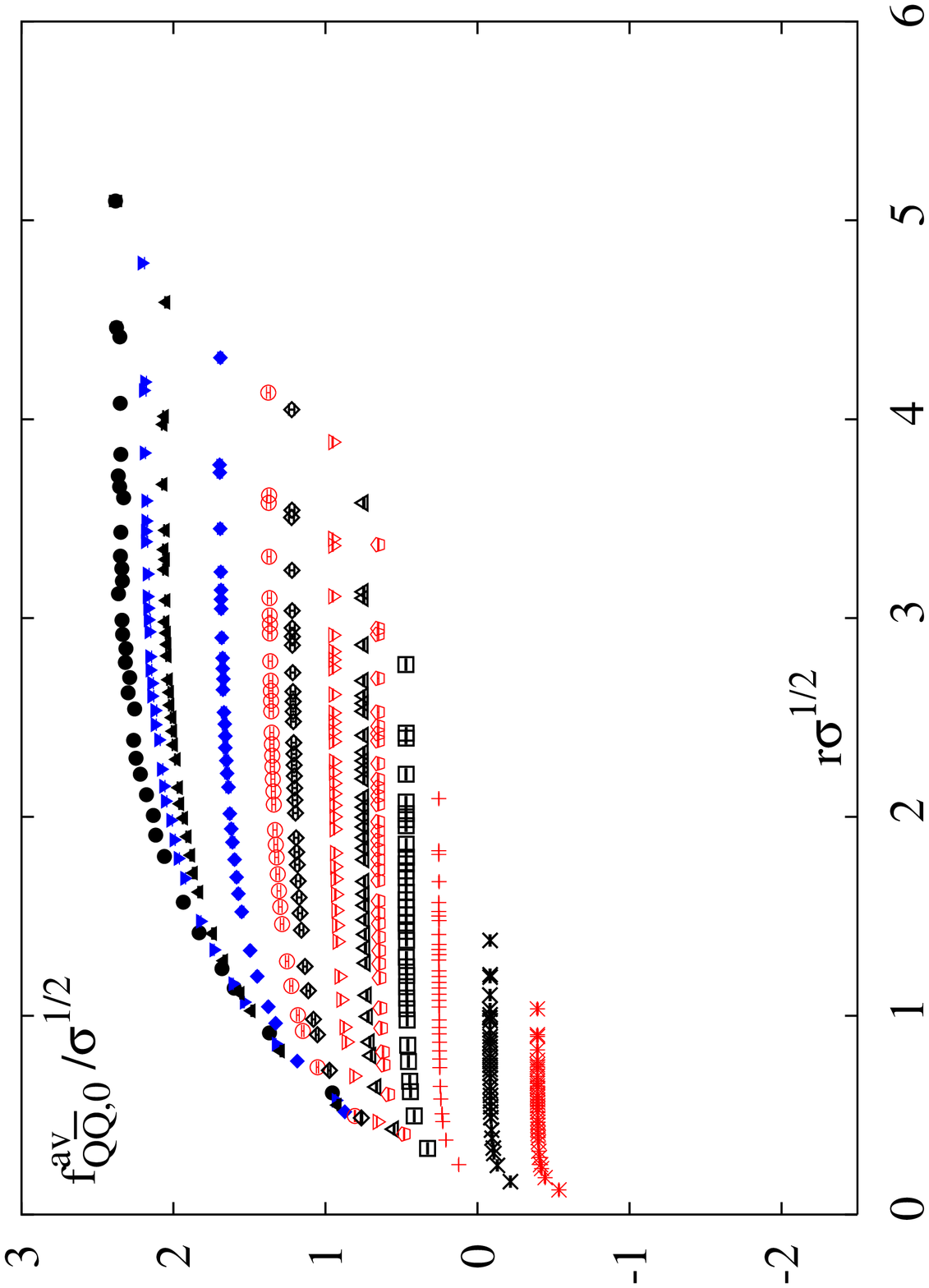} \end{turn} \\
\mbox{(a)} & \mbox{(b)}\\
\end{array}\\\eew
\caption{The $0^{th}$ order coefficients $f^1_{Q\bar{Q},0}$ and 
$f^{\rm av}_{Q\bar{Q},0}$ for the singlet and colour averaged free energy 
in the 
vicinity of $T_c$. $f^1_{Q\bar{Q},0}$ is matched to the $T=0$ heavy quark potential 
at small distances (a). \label{order0}}
\end{figure}
A purely gluonic observable ${\cal O}$ like the Polyakov loop $L(\bo{x})$ or 
a corresponding correlation function does not explicitly depend on the quark chemical potential;
it is calculated in terms of link variables $U_\rho(x)$ of the gauge
field configuration which do not explicitly depend on $\mu$. Any 
$\mu$-dependence
of the expectation value $\left< {\cal O} \right>_\mu$ thus arises from the 
$\mu$-dependence of the Boltzmann weights in the QCD partition function, 
{\it i.e.} the $\mu$-dependence of the fermion determinant. Therefore we can
apply the same method that was used for the power series expansion of the
equation of state; expanding the fermion determinant in powers of $\mu$ 
leads to a power series of our purely gluonic observable,
\be
\left< {\cal O} \right>_\mu = \left< {\cal O} \right> \cdot 
(1 + o_1 \mu + o_2 \mu^2 + \cdots)\;,
\ee
where $ \left< {\cal O} \right>$ denotes the expectation value 
of ${\cal O}$ evaluated for vanishing chemical potential.
We consider observables like the colour averaged and singlet
$Q\bar{Q}$-correlation functions,
\be
{\cal O}^{\rm av}(r) &=& 
\f{1}{{\cal N}} \, \f{1}{N_c^2} \, \suli_{\bo{x}, \bo{y}} \mbox{Tr}
 L(\bo{x}) \, \mbox{Tr} L^\dagger(\bo{y})\;, \nonumber \\
{\cal O}^{\rm 1}(r) &=& 
\f{1}{{\cal N}} \, \f{1}{N_c} \, \suli_{\bo{x}, \bo{y}} \mbox{Tr}
 L(\bo{x})  L^\dagger(\bo{y})\;, \label{corrs}
\ee
where the sum refers to all sites $\bo{x}, \bo{y}$ with $\left\| \bo{x} - \bo{y}
\right\| = r$ and ${\cal N}$ is the number of these $\bo{x}, \bo{y}$-pairs. 
As ${\cal O}^{\rm av,1}$ and the corresponding expectation values are strictly 
real for every single gauge field configuration the odd orders in the 
expansion vanish as was argued in \cite{allton}. 
For observables like the Polyakov loop itself or static quark-quark 
correlations like $\mbox{Tr} L(\bo{0})\mbox{Tr} L(\bo{r})$ we also have to take
into account the odd orders which are in general non-vanishing. In the appendix
we give explicit formulas for calculating the expansion coefficients of an
arbitrary gluonic observable up to sixth order in $\mu/T$. We have used this
to evaluate the first three, non-vanishing expansion coefficients of the 
purely real observables considered here.

We extract the colour averaged free energy of a static quark anti-quark
pair from the Polyakov loop correlation function
\be
F^{\rm av}_{Q\bar{Q}}(r, T, \mu) = -T \; \ln{
\left< {\cal O}^{\rm av}(r)
\right>_\mu}
\ee
and the colour singlet free energies from
\be
F^1_{Q\bar{Q}}(r, T, \mu) = -T \; \ln{\left< {\cal O}^{\rm 1}(r) 
\right>_\mu}\;.
\ee
We renormalize the Polyakov loop as described in \cite{kacz} such that at short
distances and vanishing chemical potential the singlet free energy
$F^1_{Q\bar{Q}}(r,T,0)$ matches the zero temperature heavy quark potential.
This also fixes the renormalization of the Polyakov loop and all its
correlation functions. In particular, this renormalizes also the colour 
averaged free energies.
As all our calculations have been performed on lattices with temporal
extent $N_\tau=4$ the smallest available distance at which this matching
could be performed is $r_0=1/4T$. 
\begin{figure}[t]
\bew
\begin{array}{cc}
\begin{turn}{270} \epsfbox{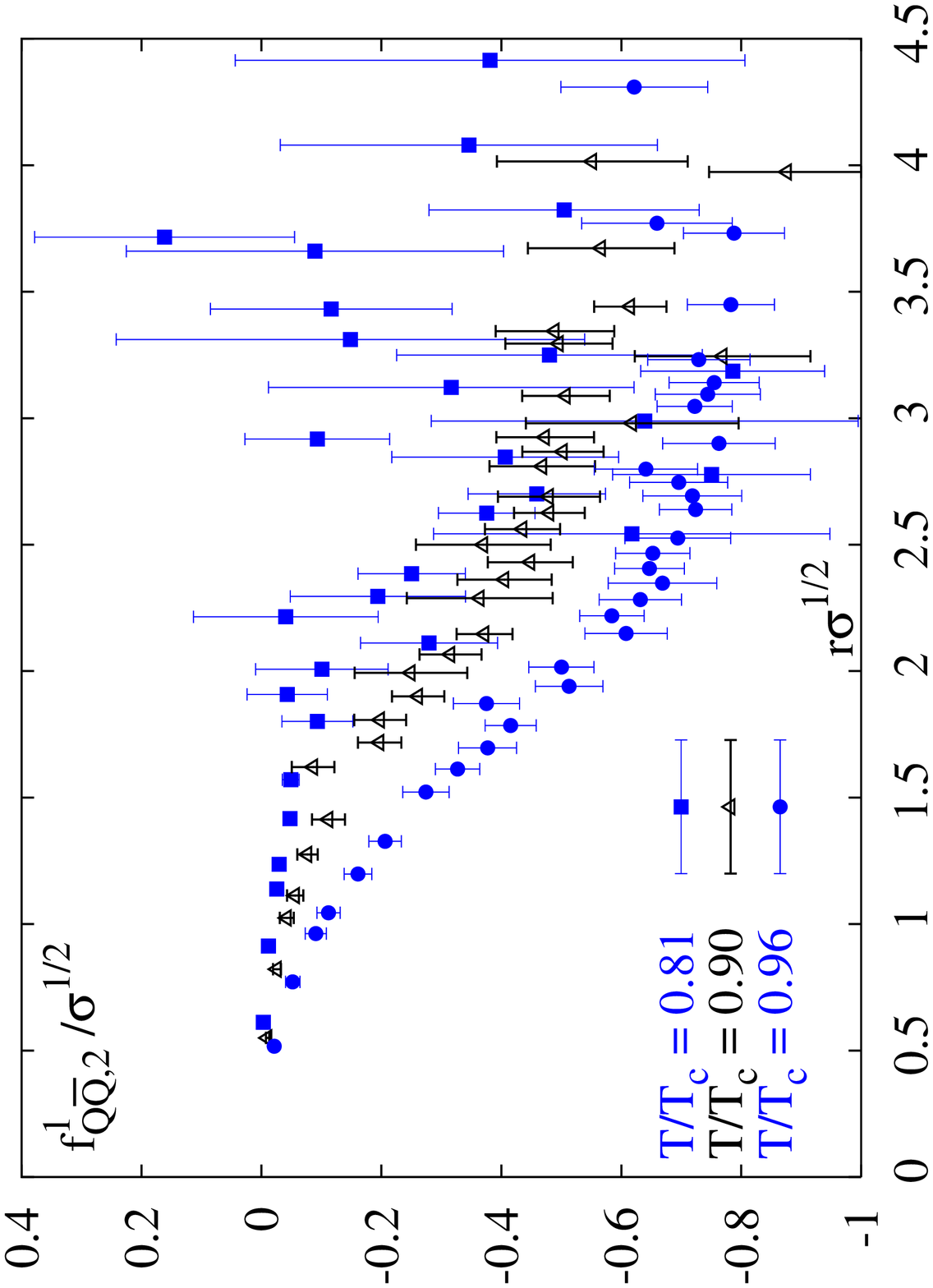} \end{turn} &
\begin{turn}{270} \epsfbox{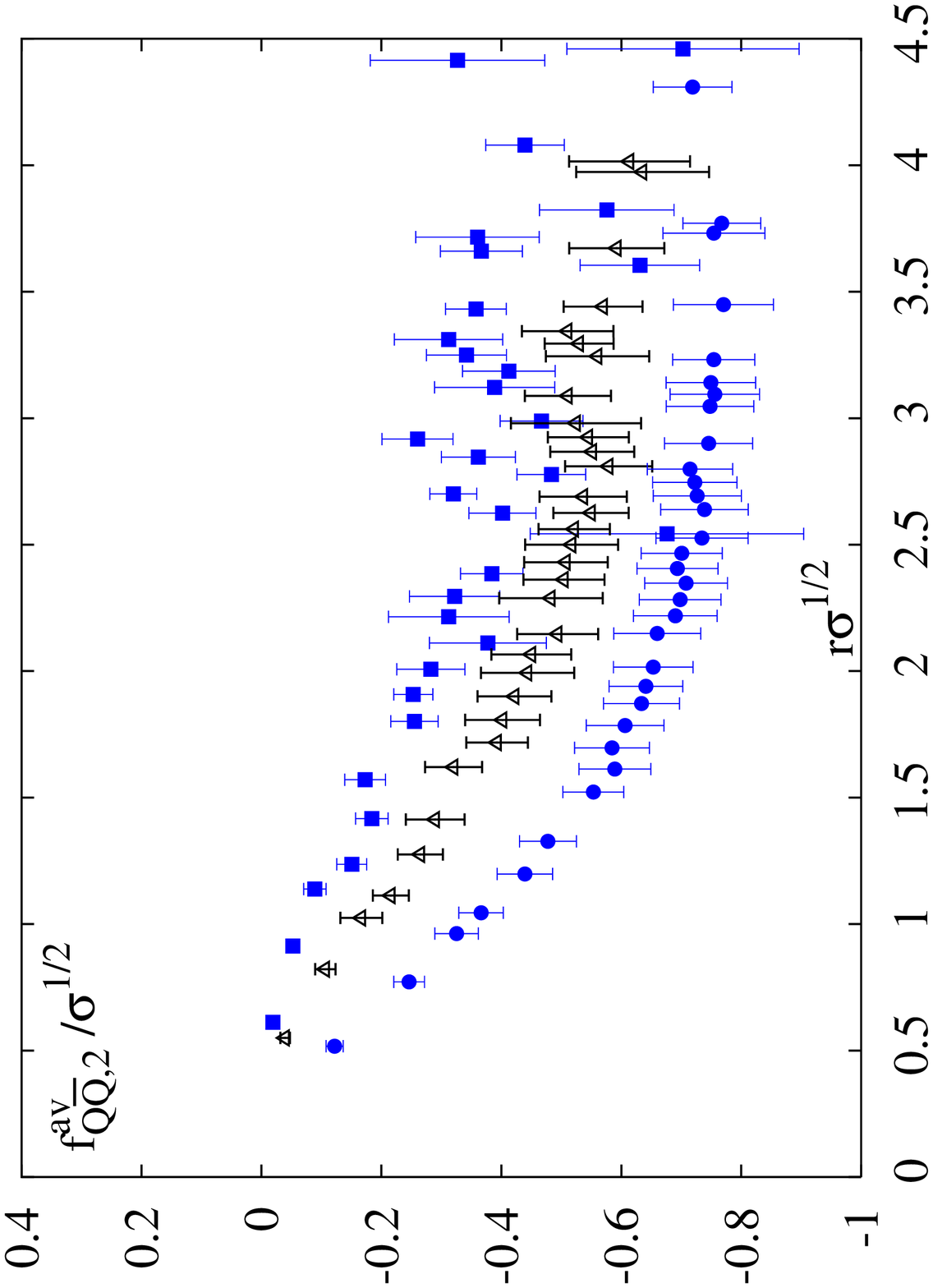} \end{turn} \\
\mbox{(a)} & \mbox{(b)}\\
\begin{turn}{270} \epsfbox{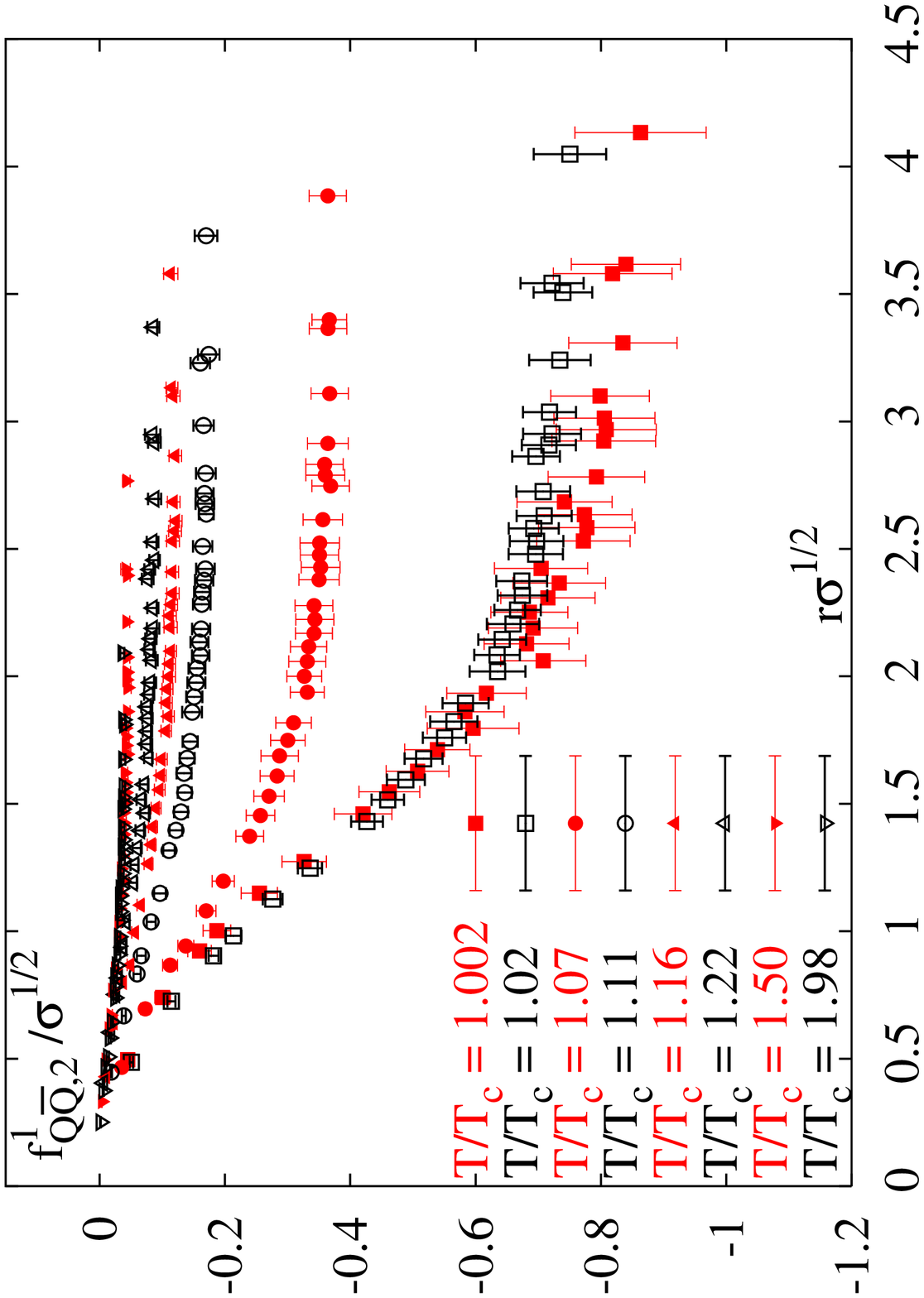} \end{turn} &
\begin{turn}{270} \epsfbox{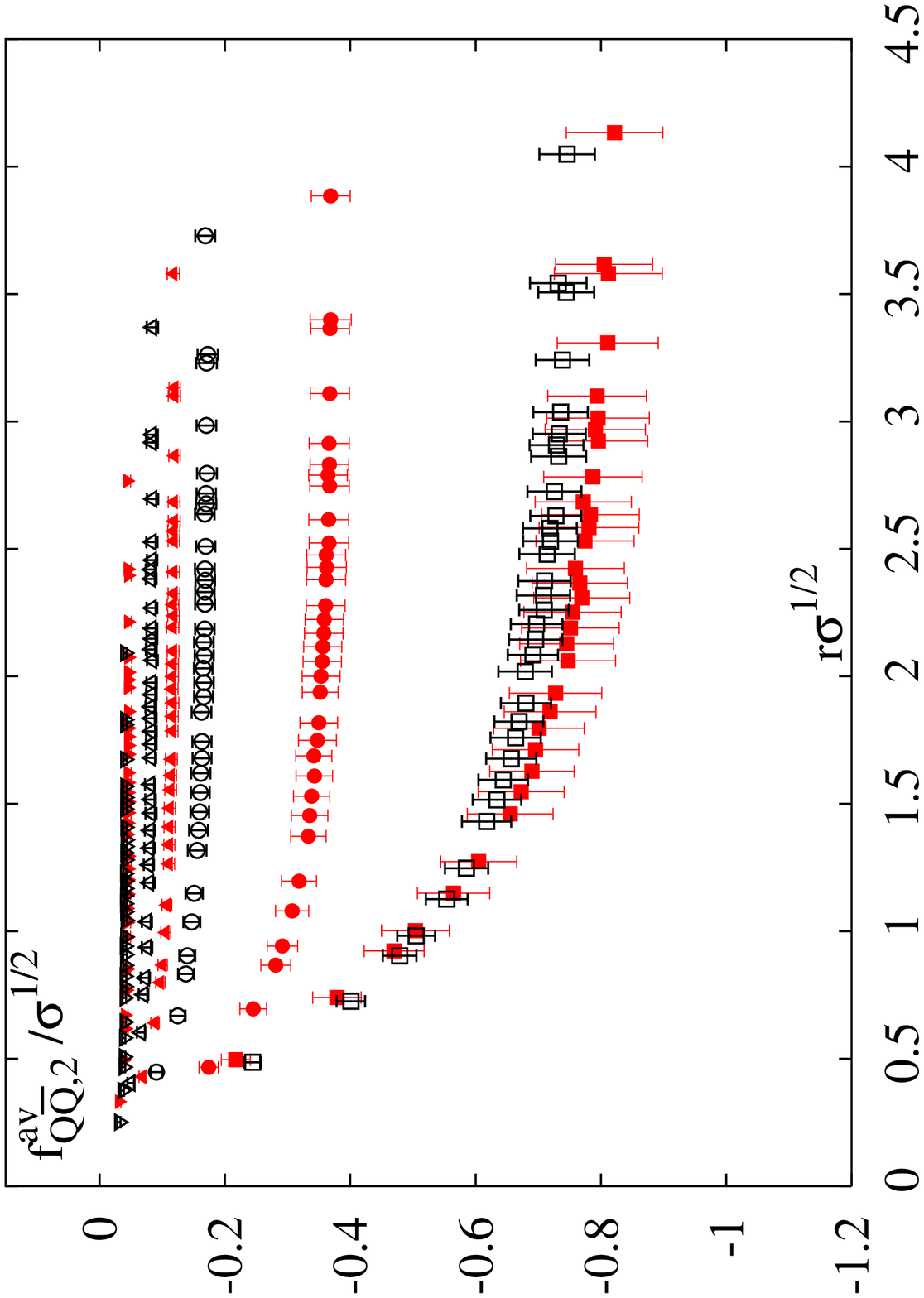} \end{turn} \\
\mbox{(c)} & \mbox{(d)}
\end{array}\\\eew
\caption{The $2^{nd}$ order coefficients of the singlet (a,c) and colour 
averaged (b,d) free energies for some selected temperatures below (a,b) 
and above (c,d) $T_c$. \label{order2}}
\end{figure}
\begin{figure}[t]
\bew
\begin{array}{cc}
\begin{turn}{270} \epsfbox{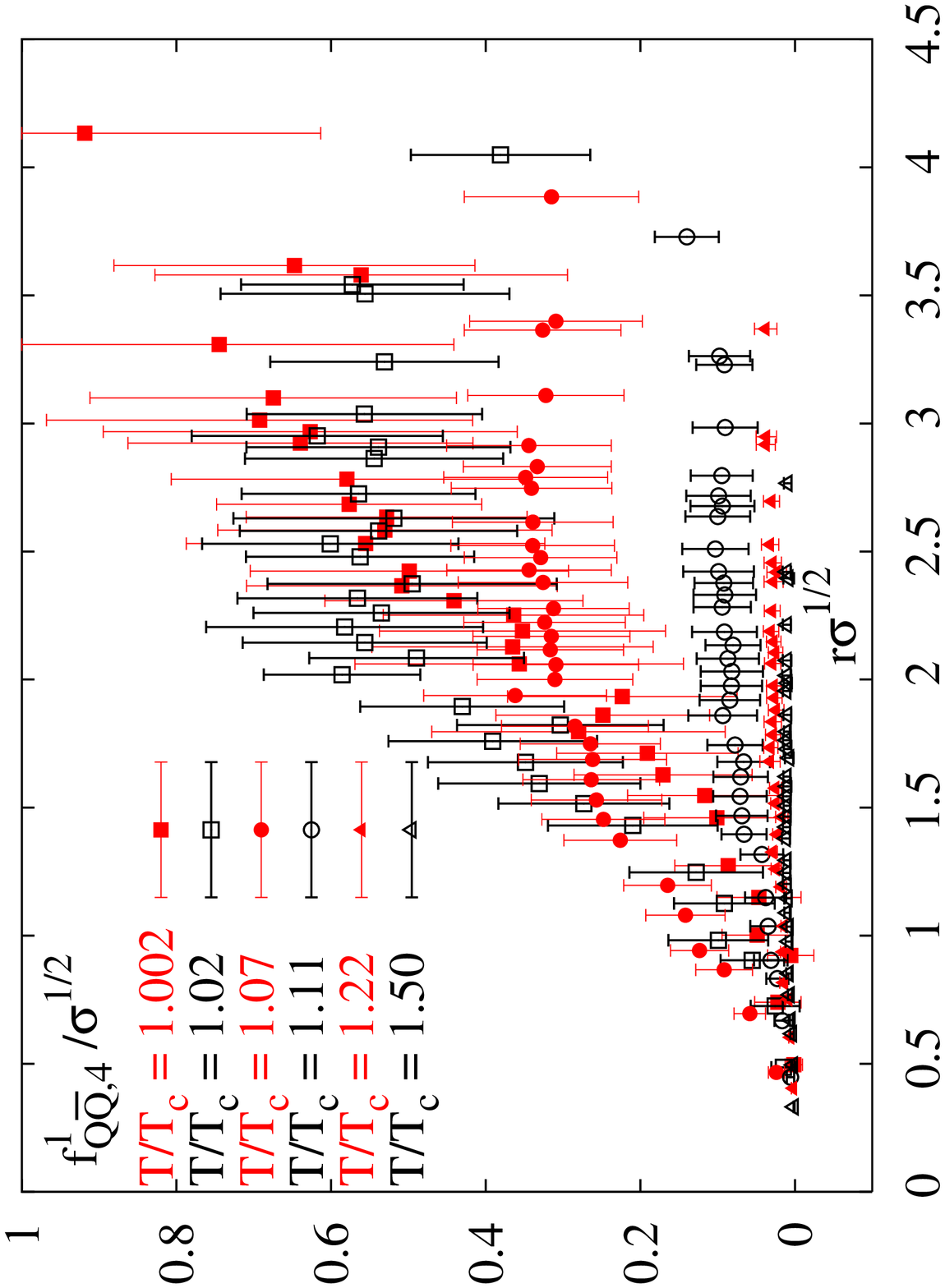} \end{turn} &
\begin{turn}{270} \epsfbox{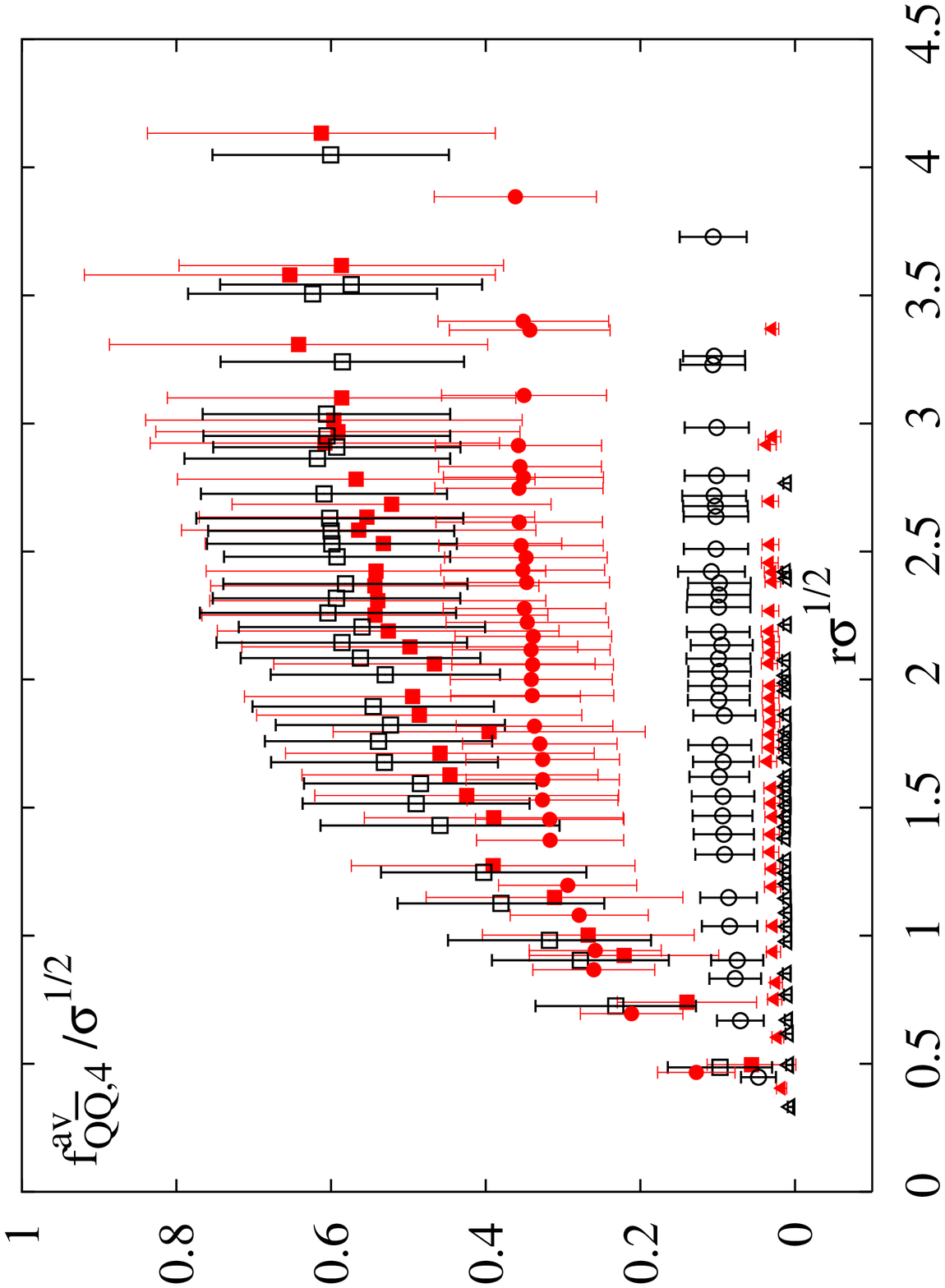} \end{turn} \\
\mbox{(a)} & \mbox{(b)}
\end{array}\\\eew
\caption{The $4^{th}$ order coefficients of the singlet (a) and colour averaged
  (b)  free energies for  temperatures above $T_c$. \label{order4}}
\end{figure}
\begin{figure}[t]
\bew
\begin{array}{cc}
\begin{turn}{270} \epsfbox{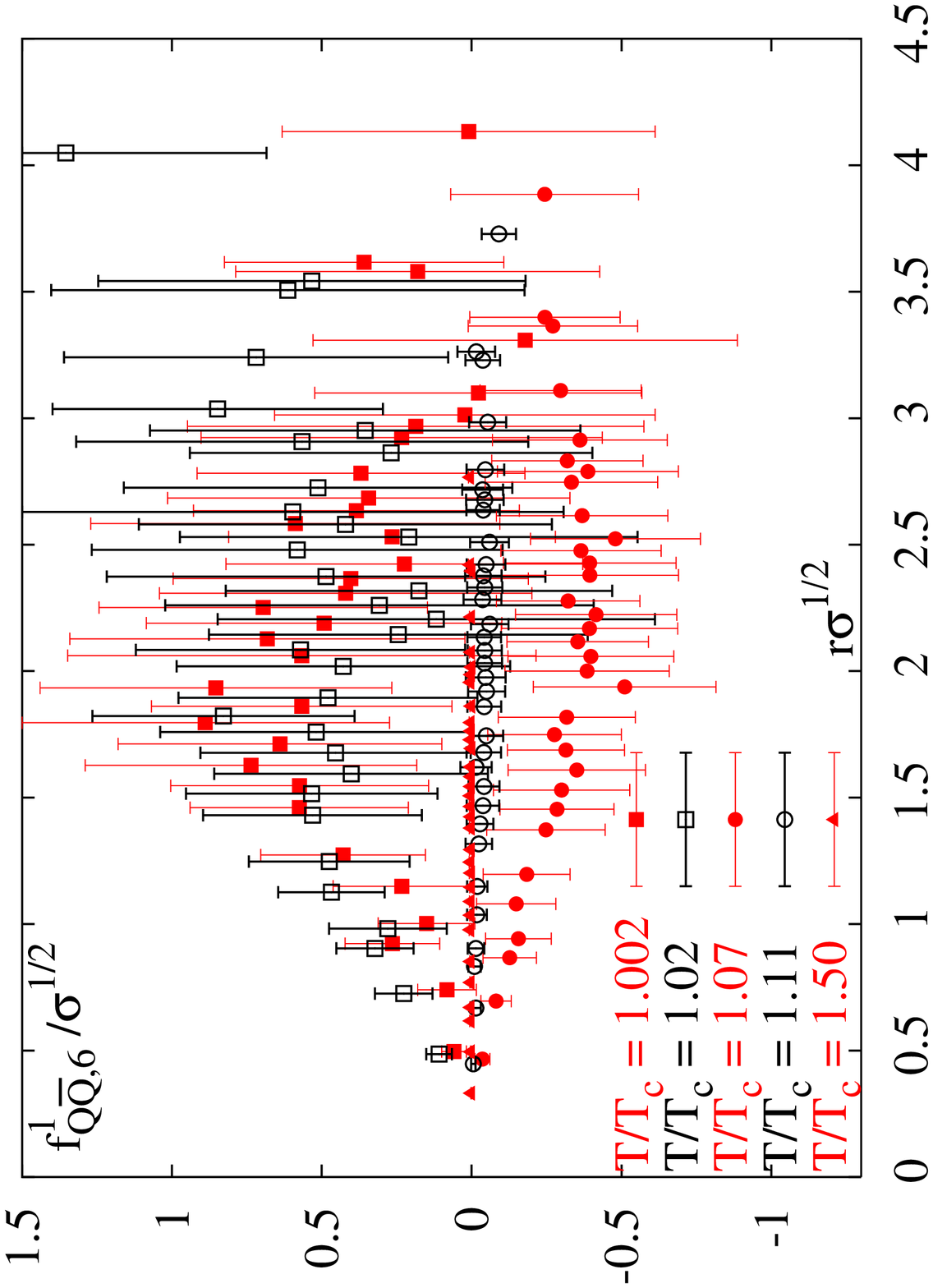} \end{turn} &
\begin{turn}{270} \epsfbox{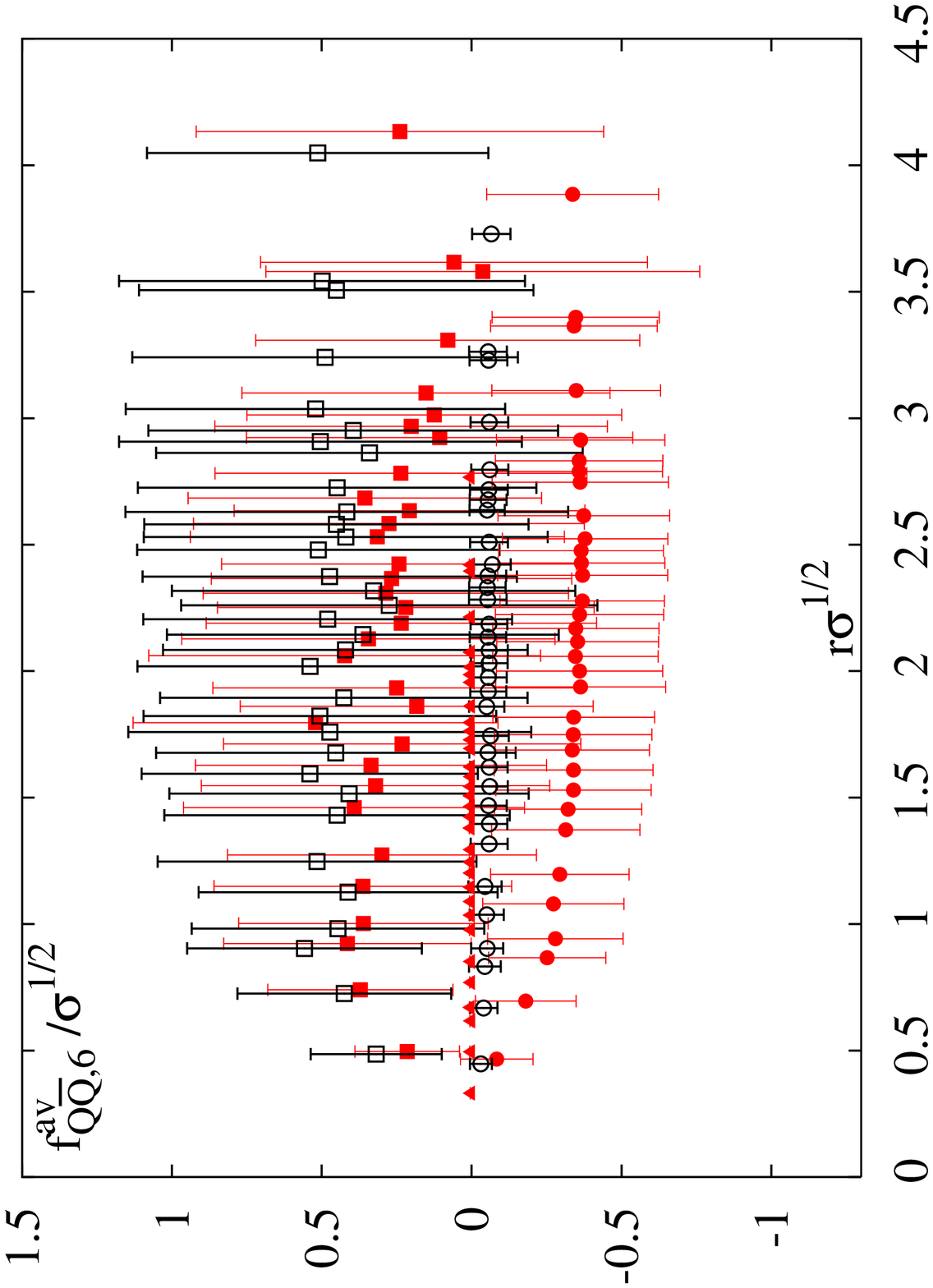} \end{turn} \\
\mbox{(a)} & \mbox{(b)}
\end{array}\\\eew
\caption{The $6^{th}$ order coefficients of the singlet (a) and colour averaged
  (b) free energies for temperatures above $T_c$. \label{order6}}
\end{figure}

In order to determine the expansion coefficients of the colour 
averaged (${\rm av}$) and singlet ($1$) free energies,
\be
F^x_{Q\bar{Q}}(r, T, \mu) = f^x_{Q\bar{Q},0}(r, T) + f^x_{Q\bar{Q},2}(r, T)
\left(\f{\mu}{T}\right)^2 + f^x_{Q\bar{Q},4}(r, T) \left(\f{\mu}{T}\right)^4 +
f^x_{Q\bar{Q},6}(r, T) \left(\f{\mu}{T}\right)^6 + {\cal
  O}\left(\left(\f{\mu}{T}\right)^8\right) \;,
\ee
with $x = {\rm av}$ and $1$, we apply (\ref{realobs}) to the corresponding
Polyakov loop correlation functions. We again note that these are strictly real
on every gauge field configuration and thus have an expansion in even powers of
$\mu/T$. Explicit formulas used for the calculation of the expansion
coefficients $f_{Q\bar{Q},n}^{\rm av}(r,T)$ and $f_{Q\bar{Q},n}^1(r,T)$ are given in the appendix.
\begin{figure}[t]
\bew
\begin{array}{cc}
\begin{turn}{270} \epsfbox{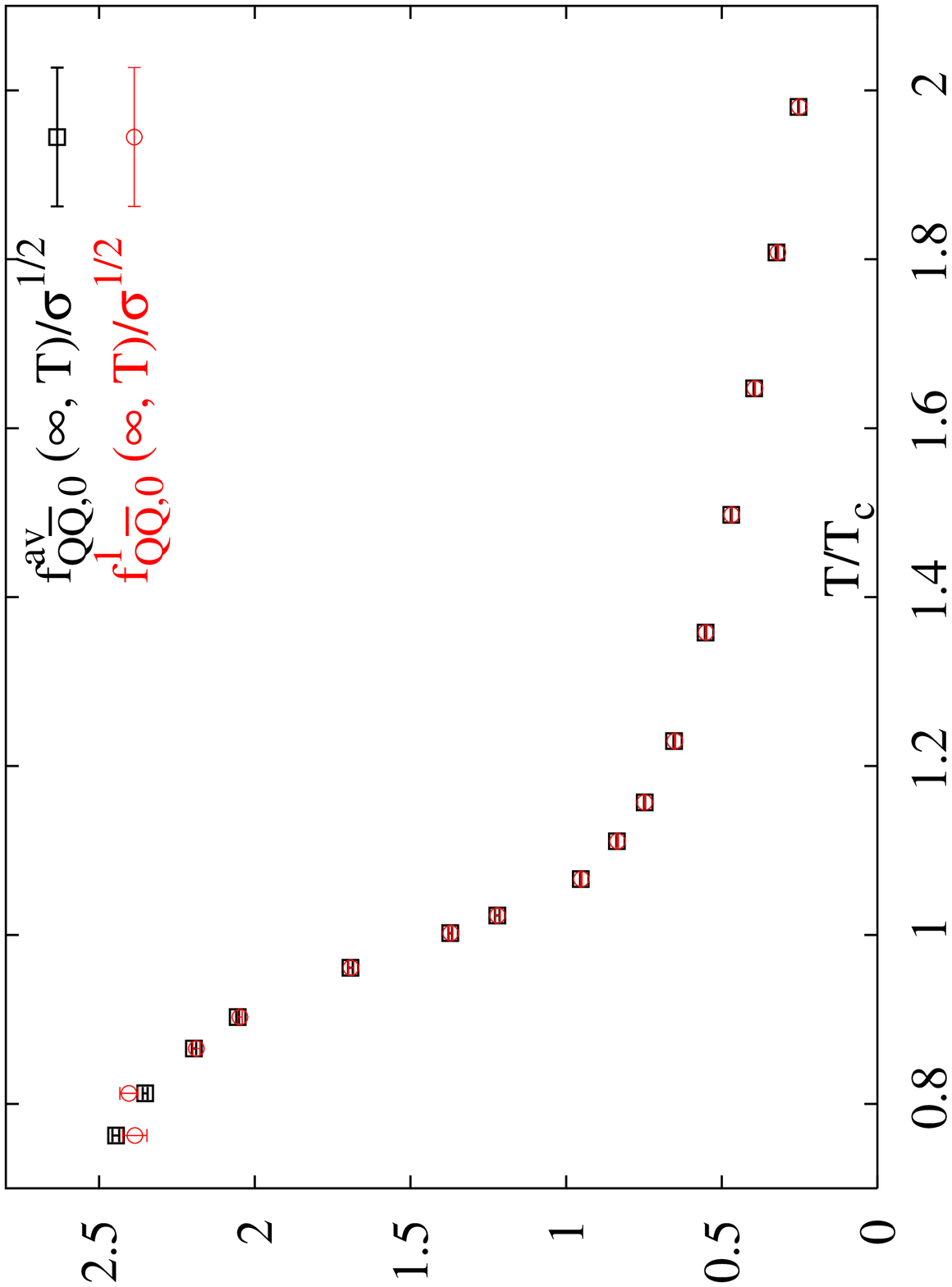} \end{turn} &
\begin{turn}{270} \epsfbox{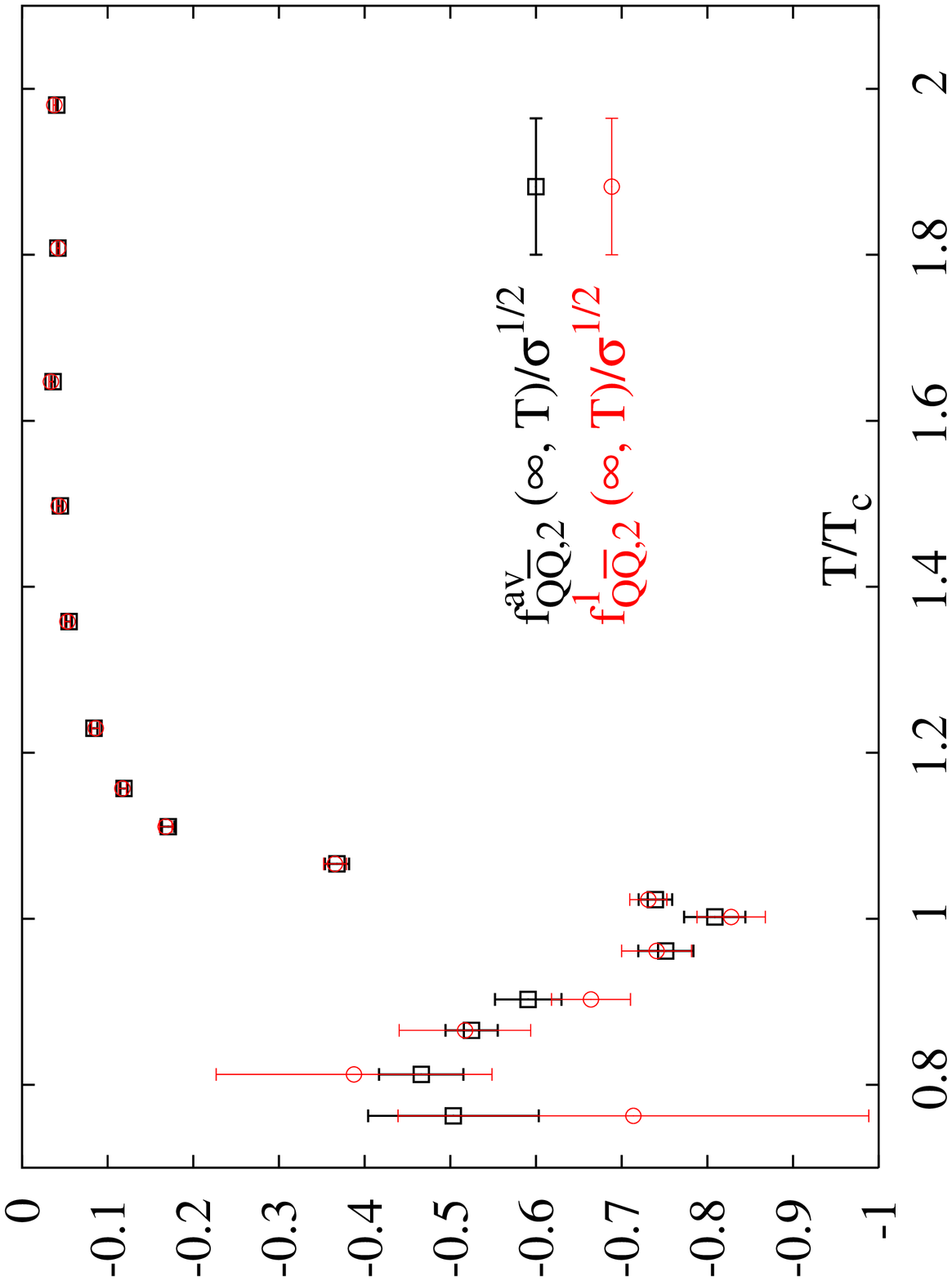} \end{turn} \\
\mbox{(a)} & \mbox{(b)}\\
\begin{turn}{270} \epsfbox{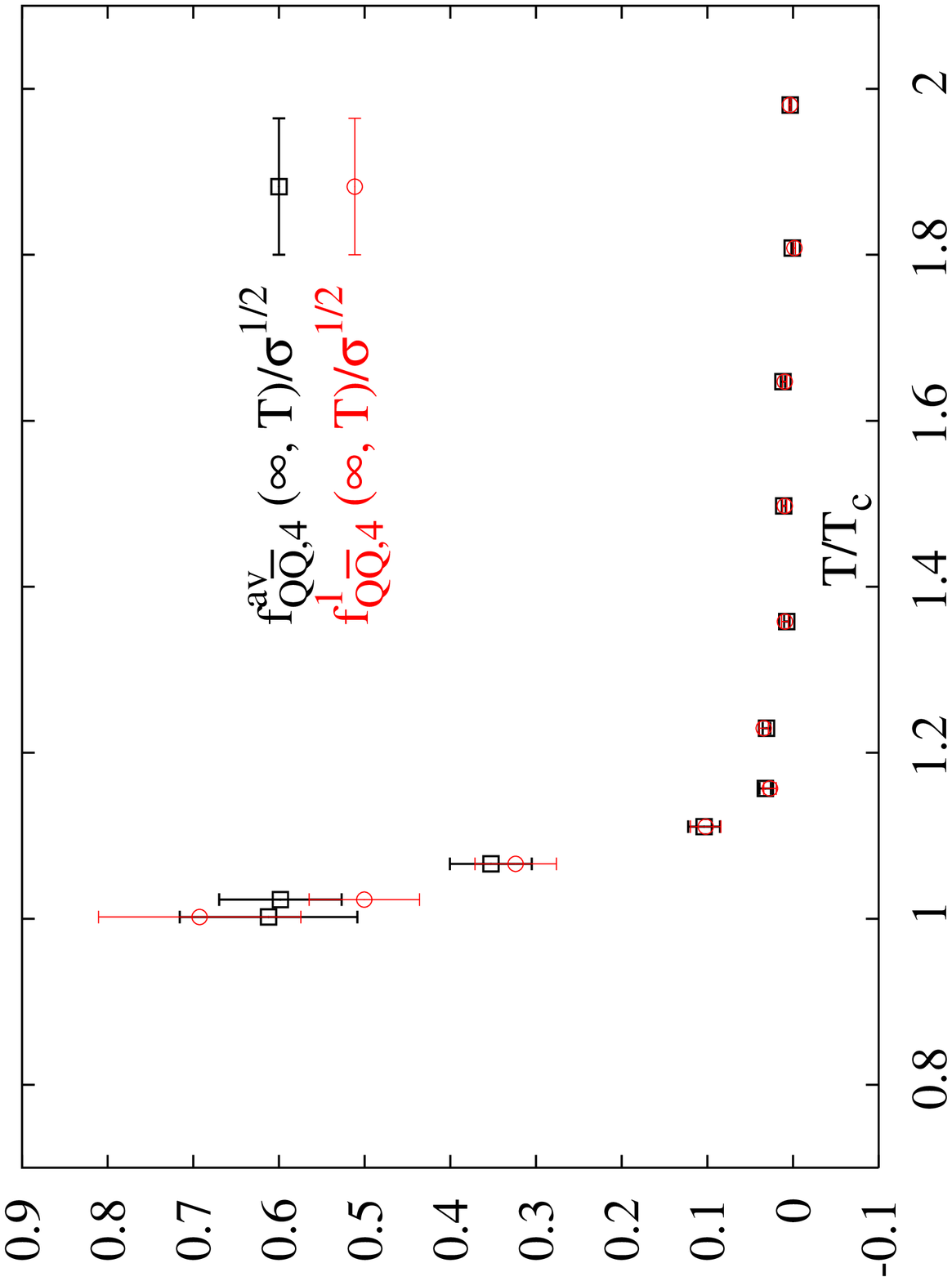} \end{turn} &
\begin{turn}{270} \epsfbox{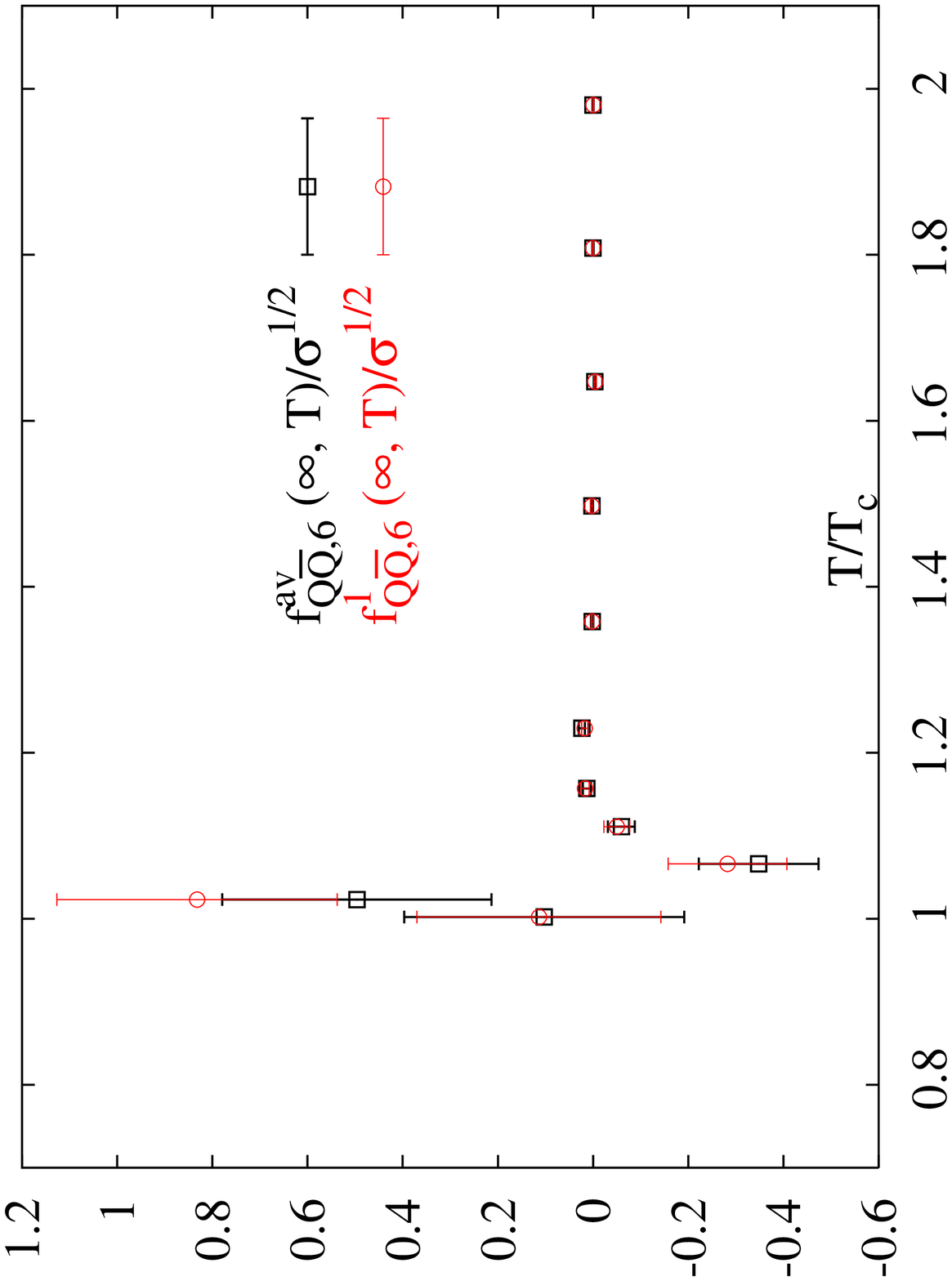} \end{turn} \\
\mbox{(c)} & \mbox{(d)}
\end{array}\eew\\
\caption{The coefficients for the singlet and colour averaged free energies 
at infinite distance $rT$ versus temperature. They have been obtained from a
weighted average of $f_{Q\bar{Q},n}^{\rm av}(r,T)$ and $f^1_{Q\bar{Q},n}(r,T)$
at the five largest distances.\label{coeffrt}}
\end{figure}
\begin{figure}[!bht]
\bew
\begin{array}{cc}
\begin{turn}{270} \epsfbox{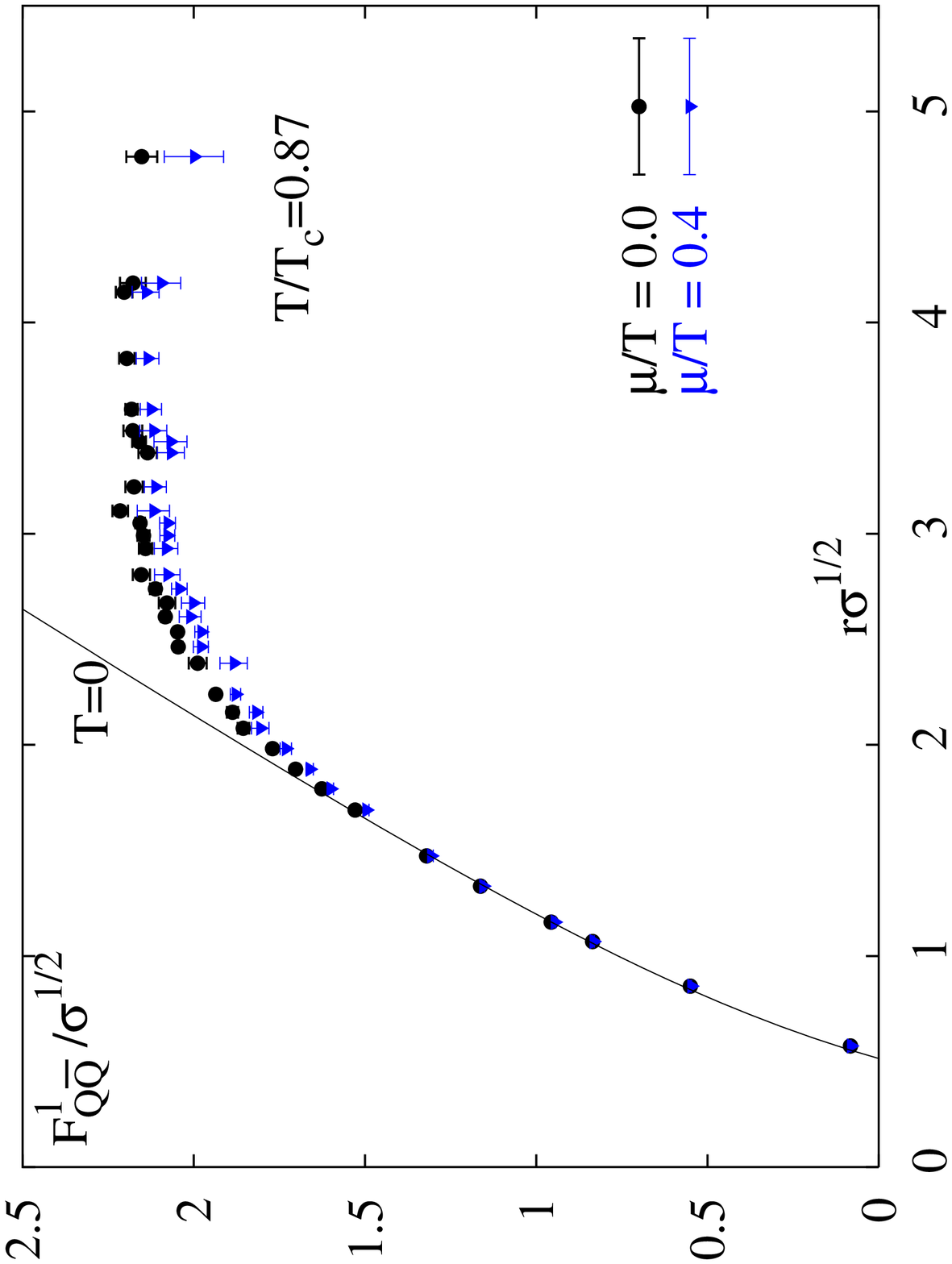} \end{turn} &
\begin{turn}{270} \epsfbox{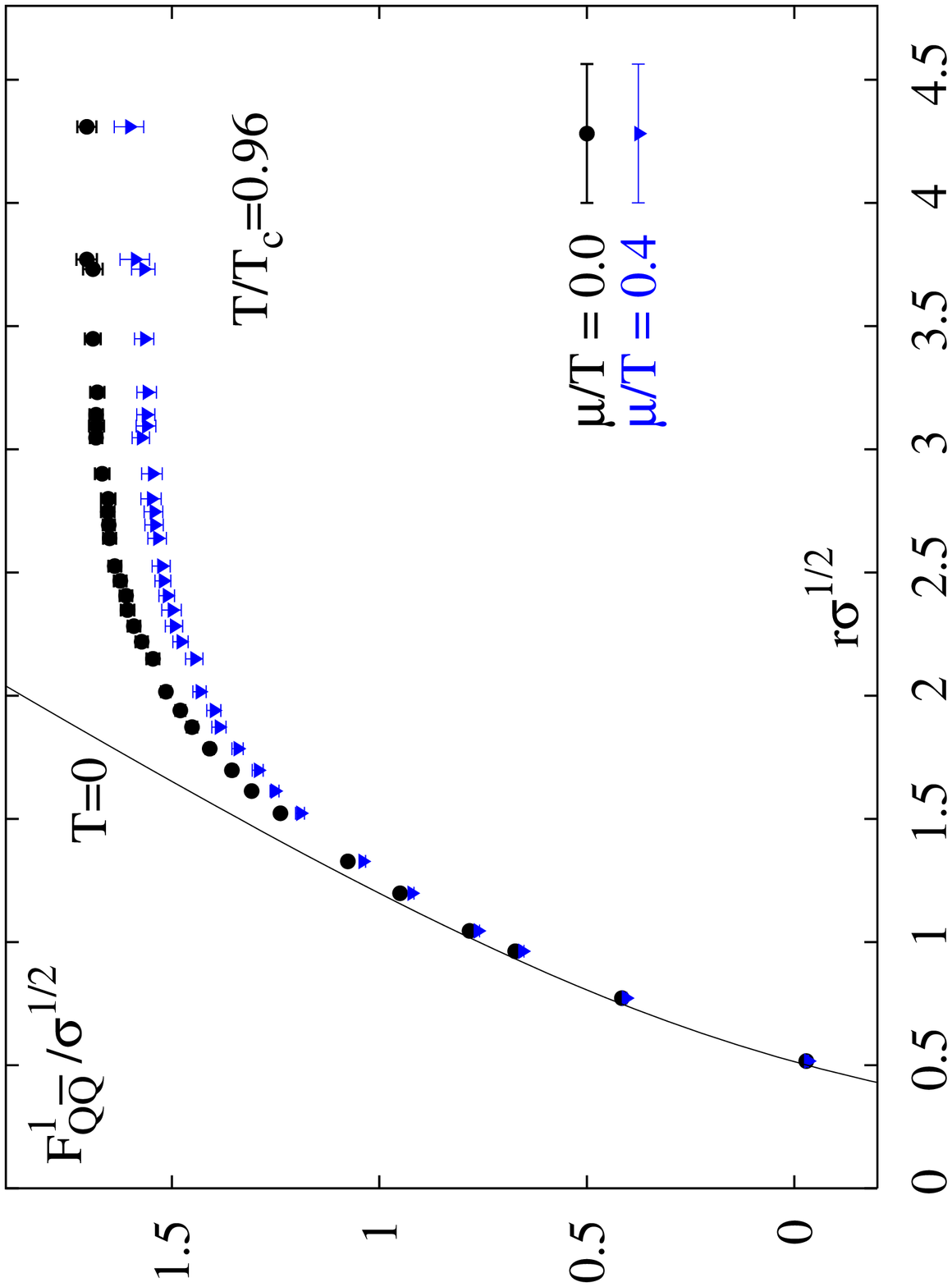} \end{turn} \\
\mbox{(a)} & \mbox{(b)}\\
\begin{turn}{270} \epsfbox{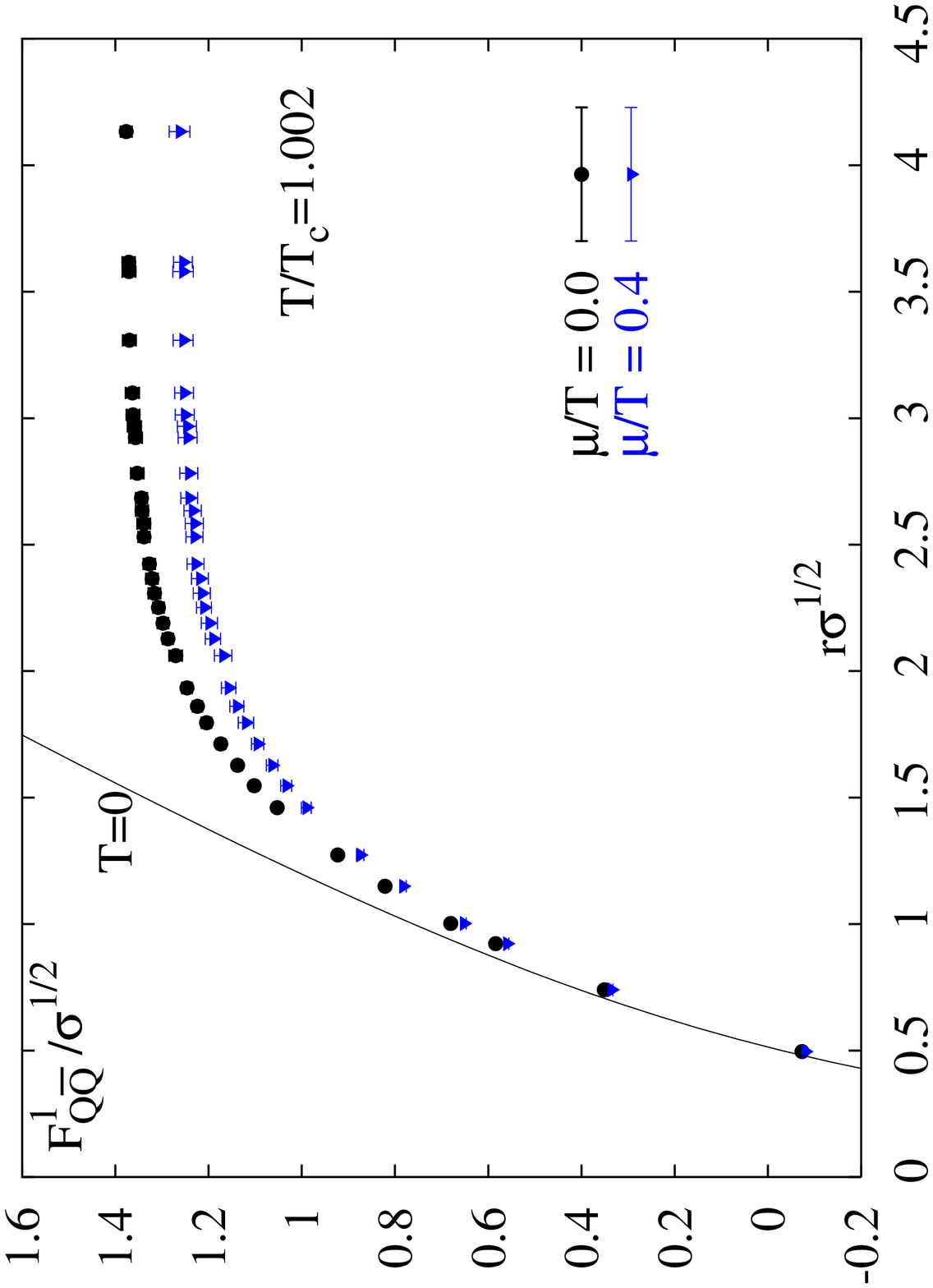} \end{turn} &
\begin{turn}{270} \epsfbox{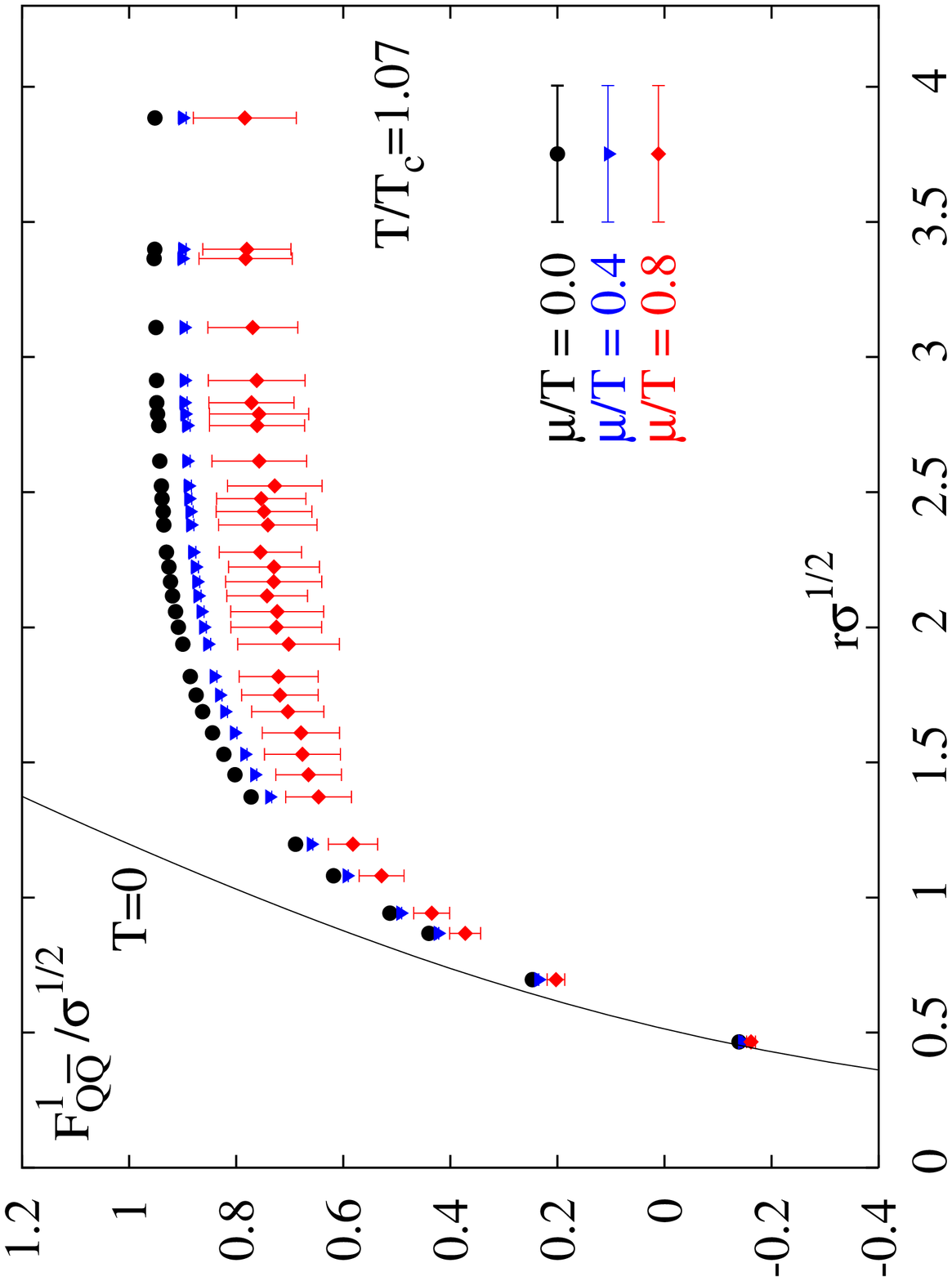} \end{turn}\\
\mbox{(c)} & \mbox{(d)}\\
\begin{turn}{270} \epsfbox{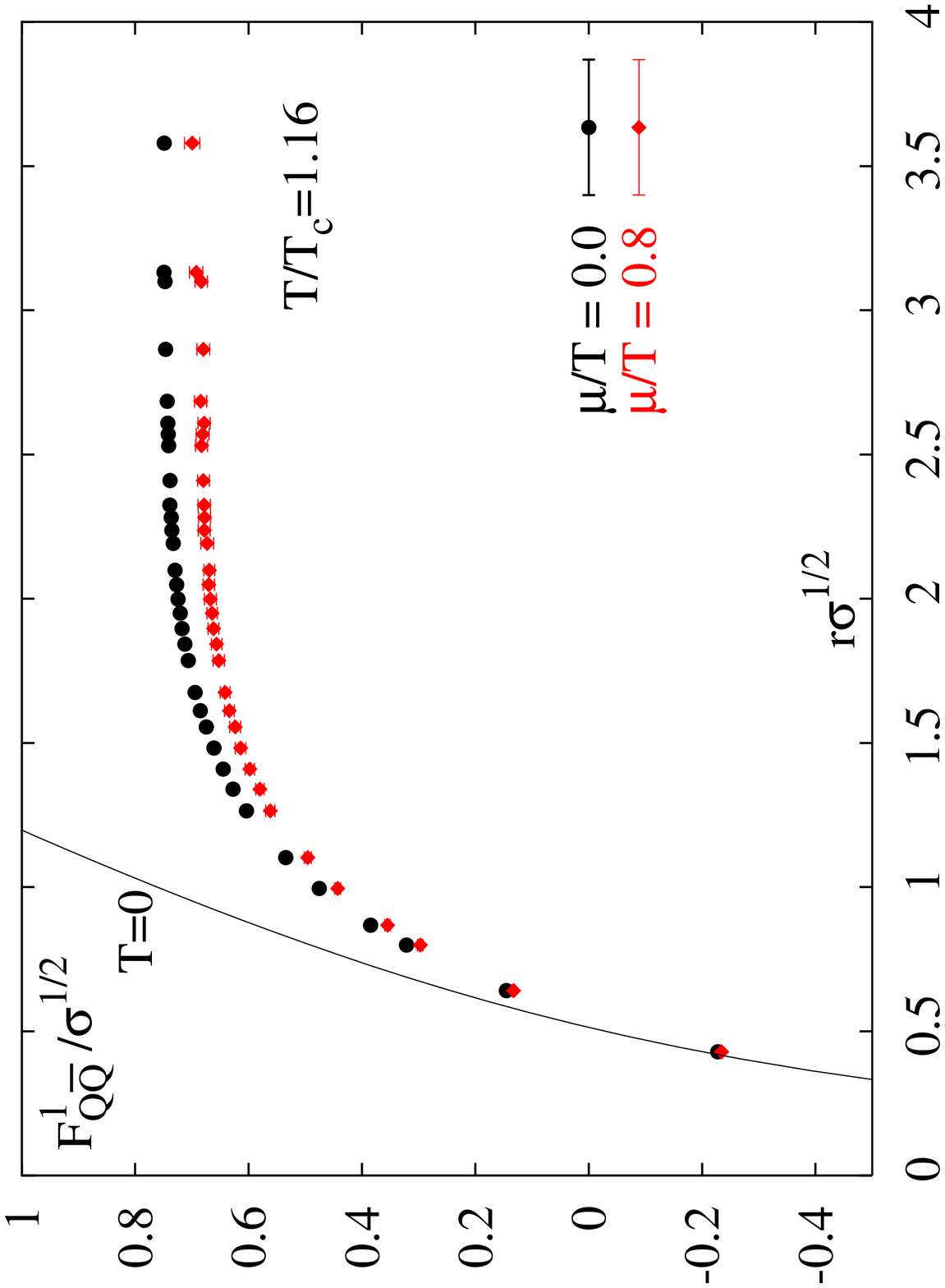} \end{turn} &
\begin{turn}{270} \epsfbox{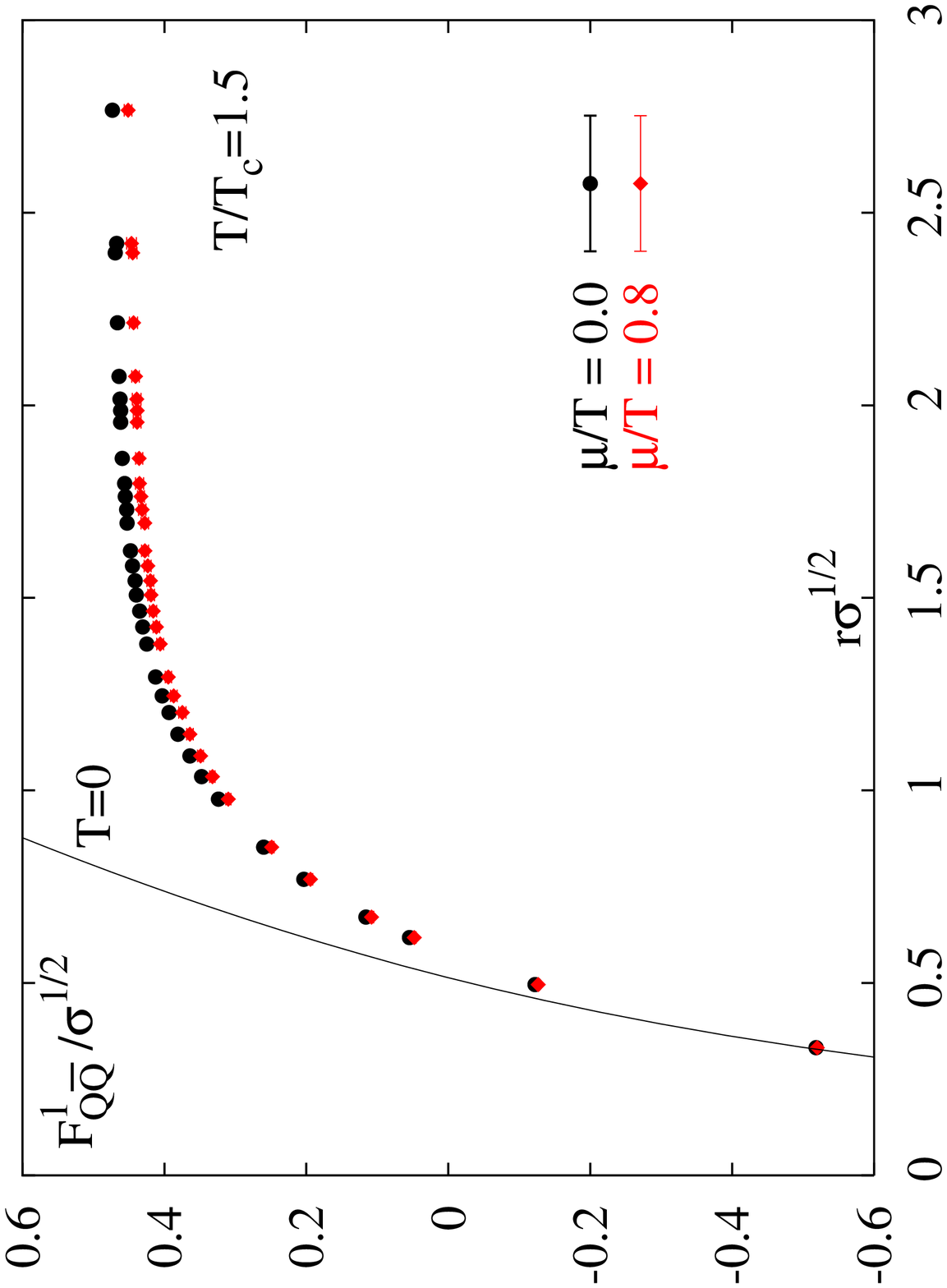} \end{turn}\\
\mbox{(e)} & \mbox{(f)}
\end{array}\\
\eew
\caption{The singlet free energies $F^1_{Q\bar{Q}}$ as function of distance
for finite chemical potential and for various temperatures. \label{singlet}}
\end{figure}

\section{Numerical results on \boldmath $Q\bar{Q}$ free energies}
\label{nume}
In Fig.~\ref{order0}-~\ref{order6} we show the leading and higher order expansion coefficients up to sixth order in $\mu/T$ expressed in units of the string tension\footnote{We use $\beta_c = 3.649$ \cite{eos,allton} as the bare pseudo-critical coupling. Results on the string tension \cite{peikert},
$a\sqrt{\sigma}$, are then used to set the temperature scale
in units of $T$.}.
We do not include data for all temperature values analyzed by us because for $T
\gg T_c$ they have very small absolute values and for $T < T_c$ they suffer from
large statistical errors and are still consistent with zero.

The leading order results, $f^{\rm av}_{Q\bar{Q},0}(r, T)$ and
$f^1_{Q\bar{Q},0}(r, T)$ are consistent with previous analyses of static quark
anti-quark free energies performed in 2-flavour QCD at $\mu = 0$ on the same data set \cite{fthq2}.
For the second order expansion coefficients we display separately results below (Fig.~\ref{order2}(a), (b)) and above (Fig.~\ref{order2}(c), (d)) the $\mu=0$ transition temperature, $T_c$.
As can be seen the second order expansion coefficients are always
negative and increase in magnitude in the vicinity of $T_c$.

The corresponding results for the $4^{\rm th}$ and $6^{\rm th}$ order expansion
coefficients are shown in Fig.~\ref{order4} and Fig.~\ref{order6}, respectively. Here we only show results above $T_c$; below $T_c$ the expansion coefficients
are consistent with being zero within errors even at rather short distances
and errors grow large for $rT \ge 1$.
We note that all expansion coefficients shown in
Figs.~\ref{order2} to \ref{order6} vanish at small distances.
This shows that a quark anti-quark pair is not affected by the
surrounding medium if its size becomes small. This observation
also justifies our procedure to renormalize the Polyakov loop by matching the
$\mu = 0$ singlet free energy to the $T=0$ heavy quark potential. The
renormalization constant is independent of $\mu$.

Also close to $T_c$, where the $\mu$-dependence of the free energies is
strongest, the absolute values of the fourth and sixth order expansion 
coefficients are of the same order as or smaller than the second order expansion 
coefficient. Therefore the 4$^{th}$ and 6$^{th}$ order contributions rapidly 
become negligible for $\mu/T < 1$.

Although the errors are large for the higher order expansion coefficients they
    show that at high temperature the $2^{nd}$ and $4^th$ order expansion 
coefficients are opposite in sign, $f^{{\rm av},1}_{Q\bar{Q},2} (r, T) < 0$ 
and  $f^{{\rm av},1}_{Q\bar{Q},4} (r, T) > 0$. 
This is consistent with the expectation
    that at high temperature the asymptotic large distance value of the heavy 
    quark free energy is proportional to the value of the Debye mass
\cite{landau}. In this limit one obtains alternating signs of the expansion
    coefficients of the heavy quark free energies when one expands the 
perturbative Debye mass \cite{debye-mu},
\be
\f{m_D(T, \mu)}{g(T)T} = \f{m_{D,0}(T)}{g(T)T}\sqrt{1 + \f{3 N_f}{(2 N_c +N_f) \pi^2} 
\left(
  \f{\mu}{T} \right)^2} \; , \label{pert}
\ee
with $m_{D,0}(T)= g(T)T\sqrt{\f{N_c}{3}+\f{N_f}{6}}$ denoting the Debye mass for
vanishing baryon chemical potential. Although the statistical significance
of our results for $f^{{\rm av},1}_{Q\bar{Q},6} (r, T)$ rapidly drops with 
increasing temperature this pattern of alternating signs seems to be valid 
also at sixth order at least for temperatures $T\gsim 1.05 T_c$.

Except for temperatures close to the transition temperature the
asymptotic behaviour of the free energies is reached at distances 
$rT\gsim 1.5$. We determined their large distance value by taking the weighted average of the values at the five largest distances. 
The results are shown in
Fig.~\ref{coeffrt}. We note that $|f^{{\rm av},1}_{Q\bar{Q},2}(\infty, T)|$ 
have a pronounced peak at $T_c$. This also holds for $|f^{{\rm
    av},1}_{Q\bar{Q},2}(r, T)|$ evaluated at any fixed distance $r$. In fact, $f^{{\rm av},1}_{Q\bar{Q},2}(r, T)$ is proportional to the second derivative of
a partition function including a pair of static sources, $Q\bar{Q}$. It thus
shows the characteristic properties of a susceptibility in the vicinity of a (phase) transition point.

Fig.~\ref{coeffrt} also shows that at large distances, within the statistical
errors of our analysis, the expansion coefficients for the colour averaged and singlet free energies approach identical values,
\be
f^{\rm av}_{Q\bar{Q},n} (\infty, T) &=& f^1_{Q\bar{Q},n} (\infty, T) \; ,\label{finfty}
\ee
where
\be
f_{Q\bar{Q},n}^x (\infty,T) &=& \lim_{r\rightarrow \infty} 
f^x_{Q\bar{Q},n}(r, T)\; ,\quad {\rm with}\; x={\rm av},~1 \; .
\ee
This has been noted before at $\mu = 0$ and suggests that at large
distances, e.g. for $rT \gsim 1.5$, the quark anti-quark sources are 
screened independently from each other; their relative colour orientation thus
becomes irrelevant.

Including all terms up to sixth order we calculated the singlet and colour
averaged free energies in the range
from $\mu/T = 0.0$ up to $0.8$. Results for the colour
singlet free energies evaluated at a few values of temperature are shown in
Fig.~\ref{singlet}. Similar results hold for the colour averaged free energies.
The free energies decrease relative to their values at $\mu/T=0$ for all 
temperatures above and below $T_c$.
At small distances the curves always agree within errors. With
increasing distance a gap opens up which reflects the decrease in free
energy at non zero $\mu$. 
As indicated by the asymptotic values $f^{{\rm av},1}_{Q\bar{Q},2}(\infty, T)$,
which give the dominant $\mu$-dependent contribution at large distances, the
medium effects are largest close to the transition temperature and 
become smaller with increasing temperature.

\section{Screening masses}

For temperatures above $T_c$ and large distances $r$ the heavy quark
free energies are expected to be screened,
\label{scre}

\begin{figure}[t]
\bew
\begin{array}{cc}
\begin{turn}{270} \epsfbox{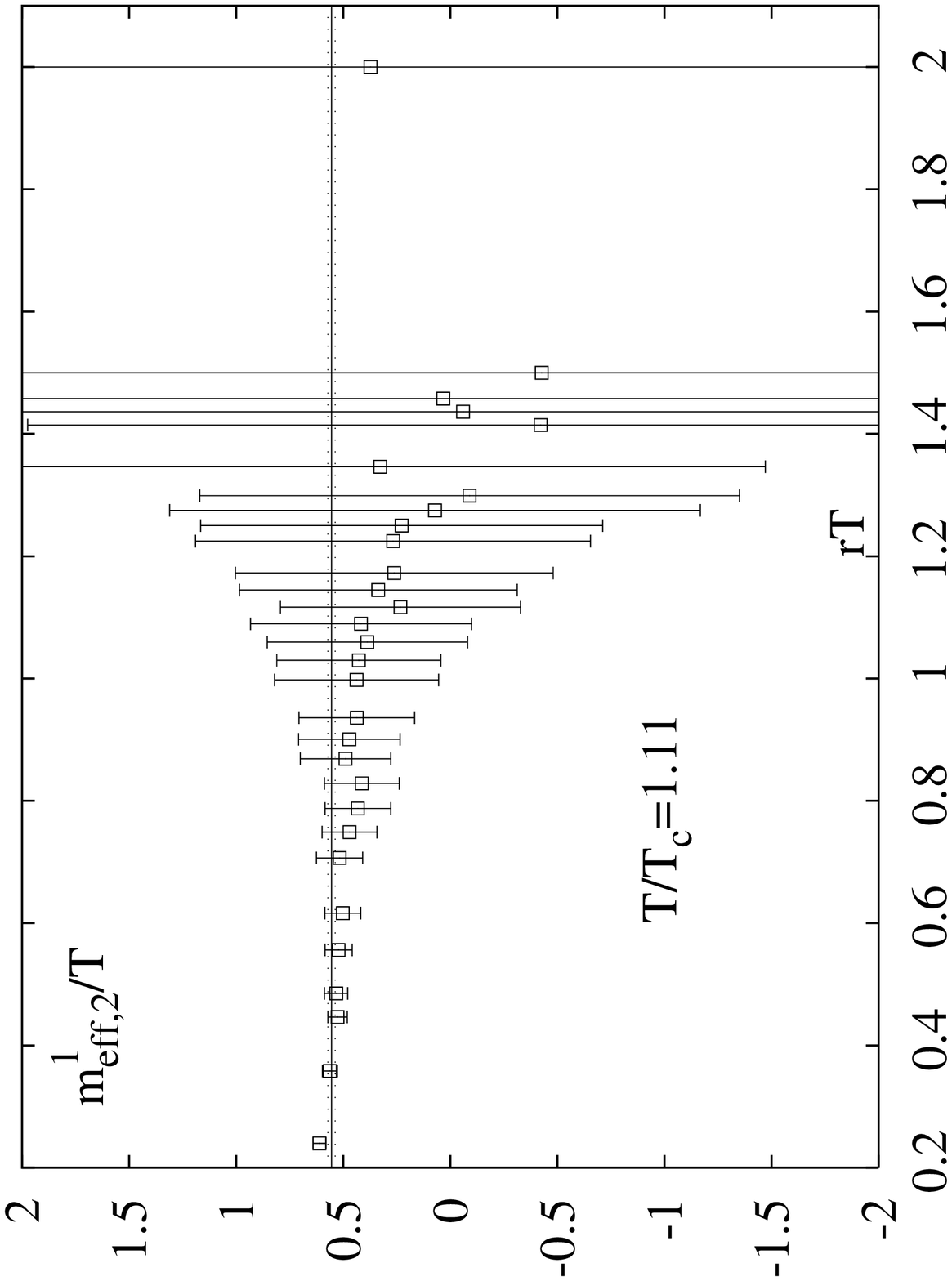} \end{turn} &
\begin{turn}{270} \epsfbox{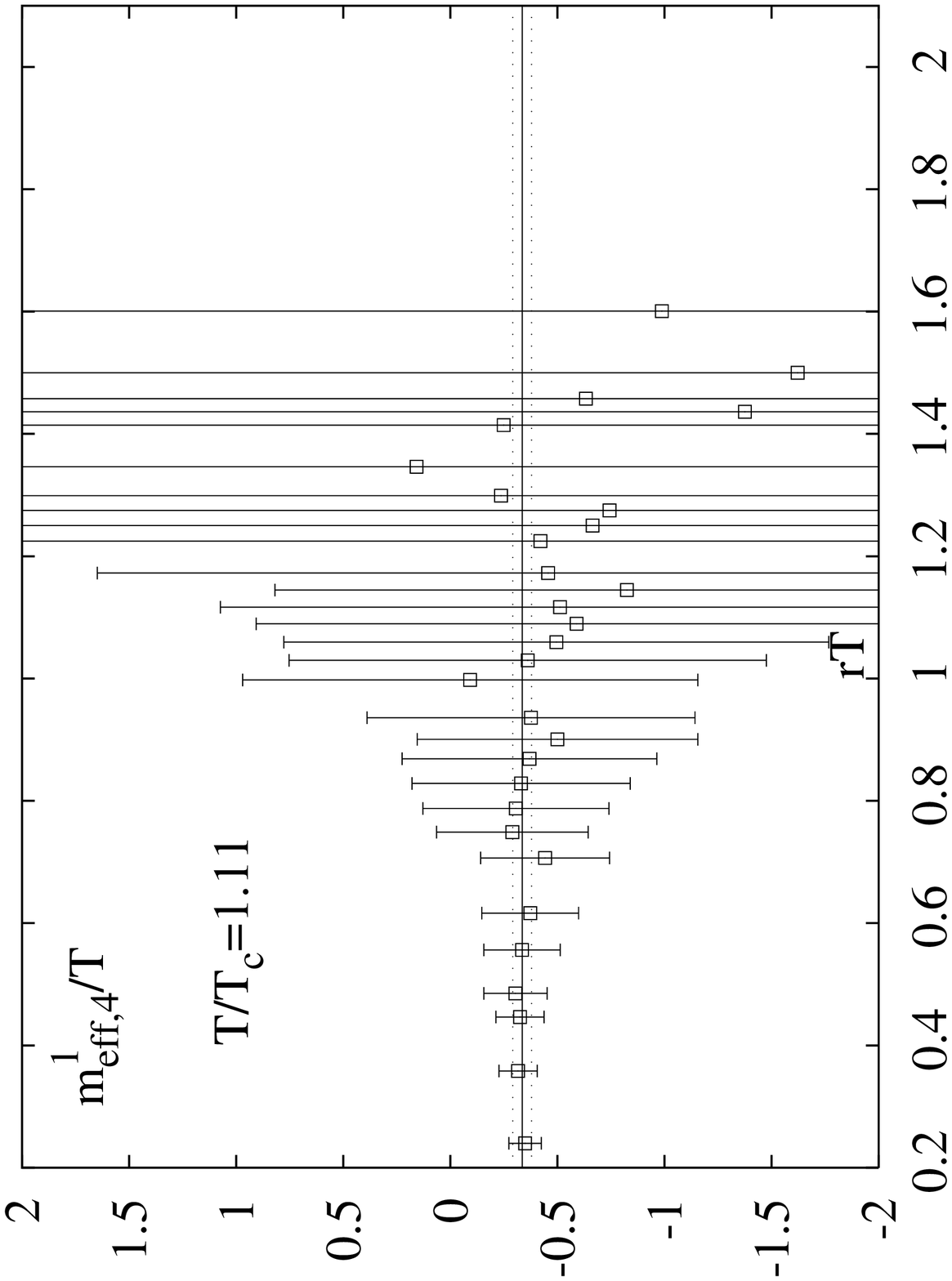} \end{turn}\\
\mbox{(a)} & \mbox{(b)}\\
\begin{turn}{270} \epsfbox{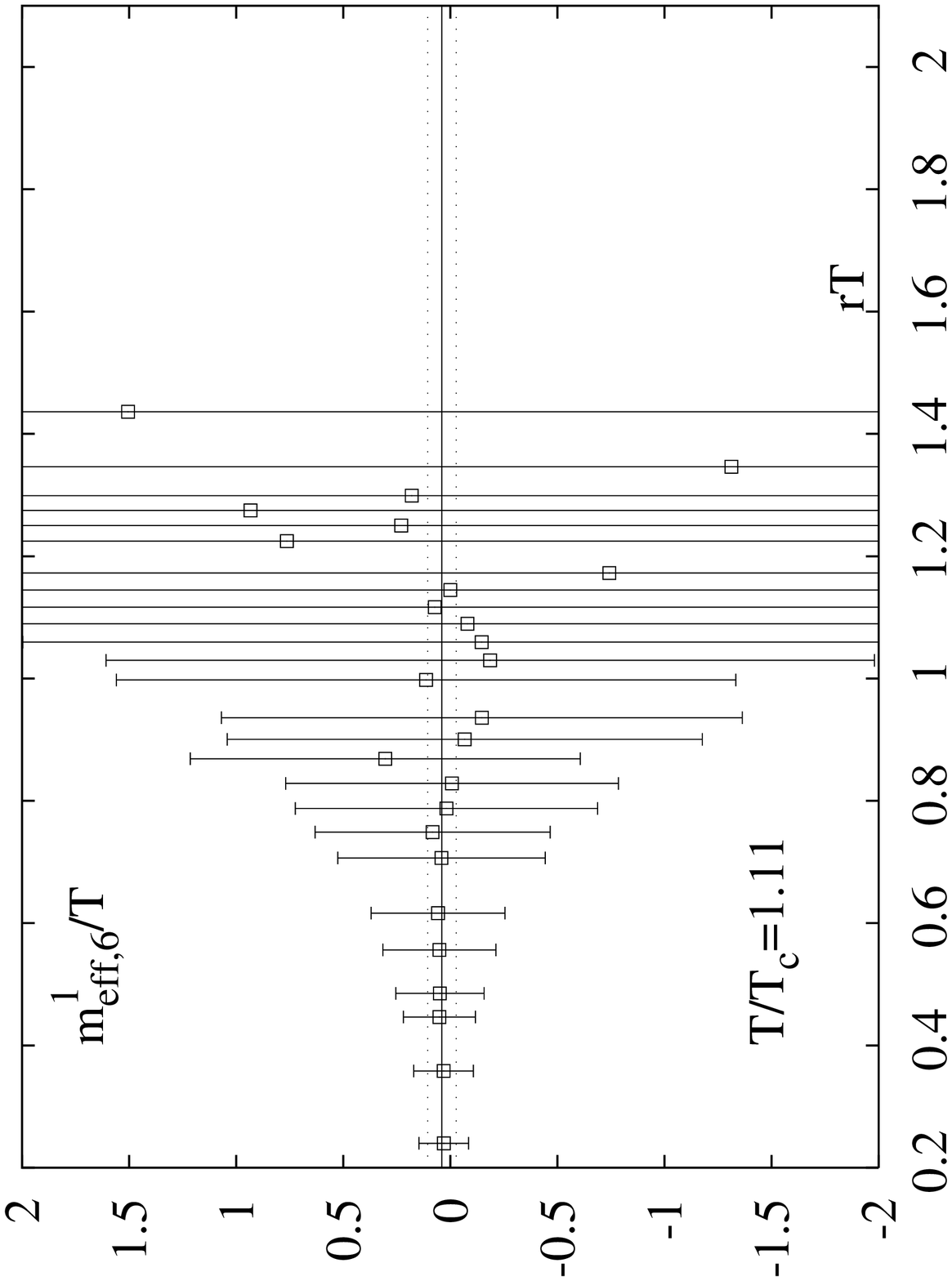} \end{turn} &\\
\mbox{(c)} &
\end{array}\\
\eew
\caption{Example for the distance dependent effective masses converging
at large distance to the expansion coefficients for the screening 
mass which are represented by the horizontal lines. \label{df2df0}}
\end{figure}
\begin{figure}[t]
\bew
\begin{array}{cc}
\begin{turn}{270} \epsfbox{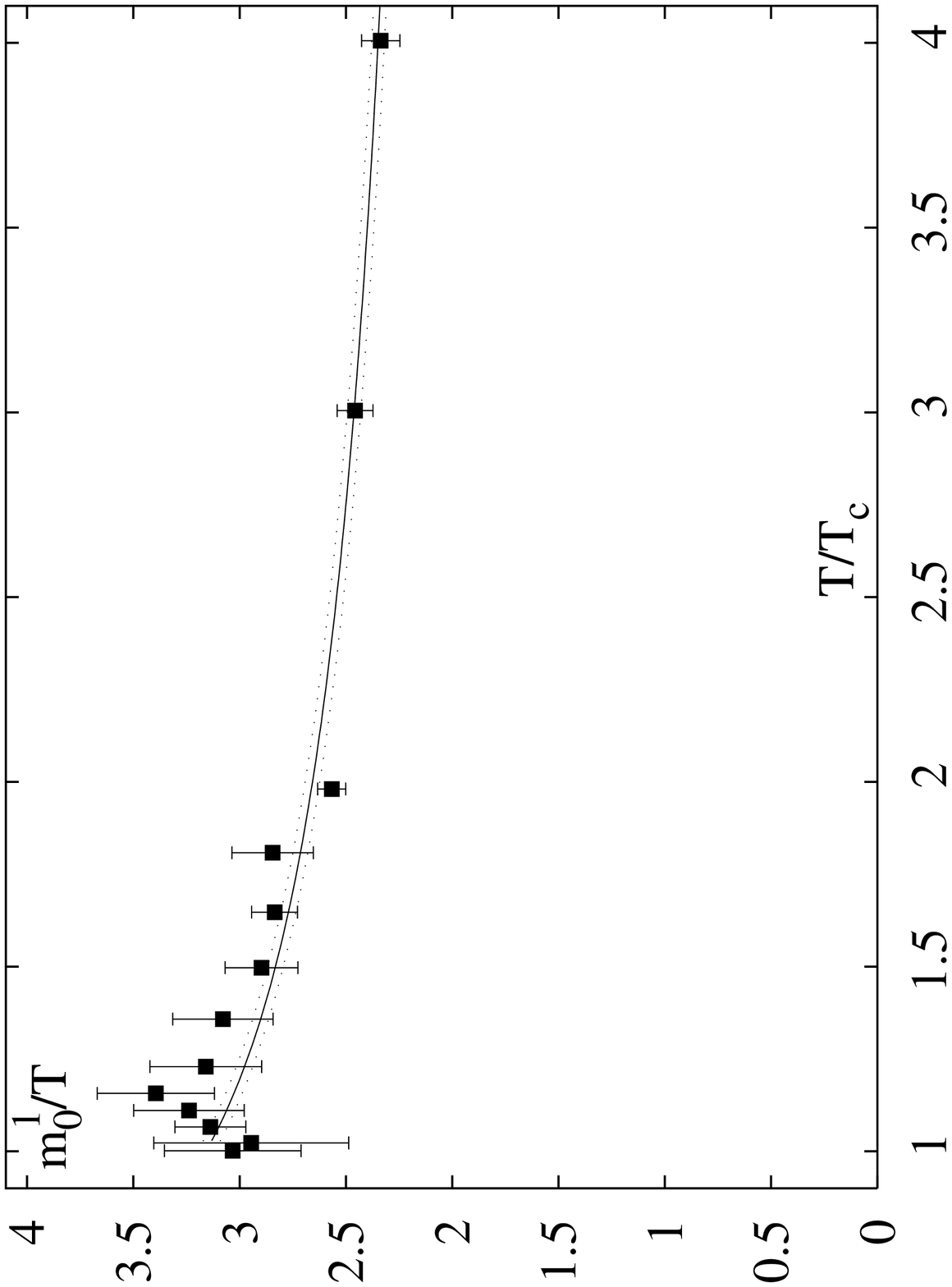}   \end{turn} &
\begin{turn}{270} \epsfbox{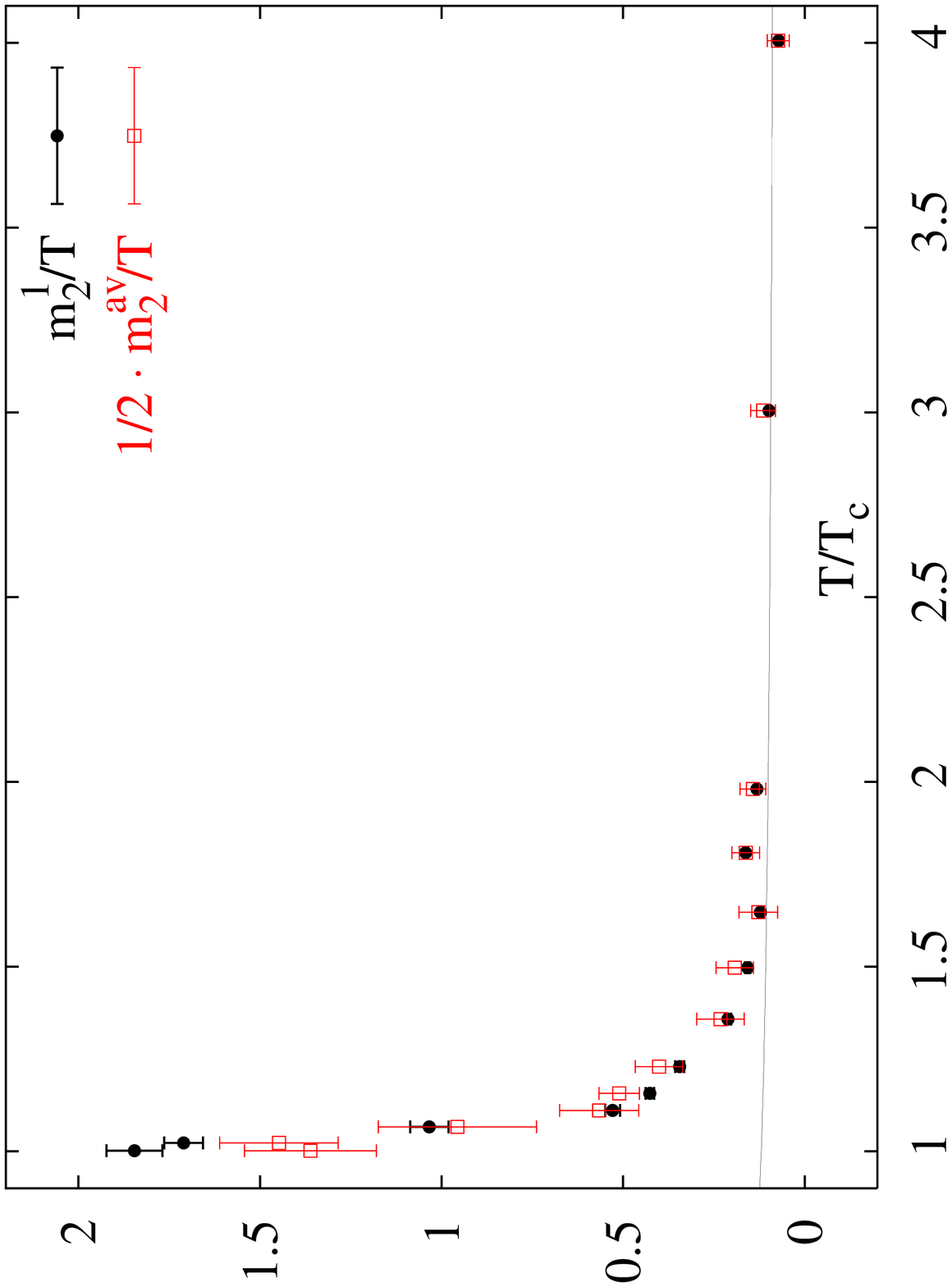} \end{turn}\\
\mbox{(a)} & \mbox{(b)}\\
\begin{turn}{270} \epsfbox{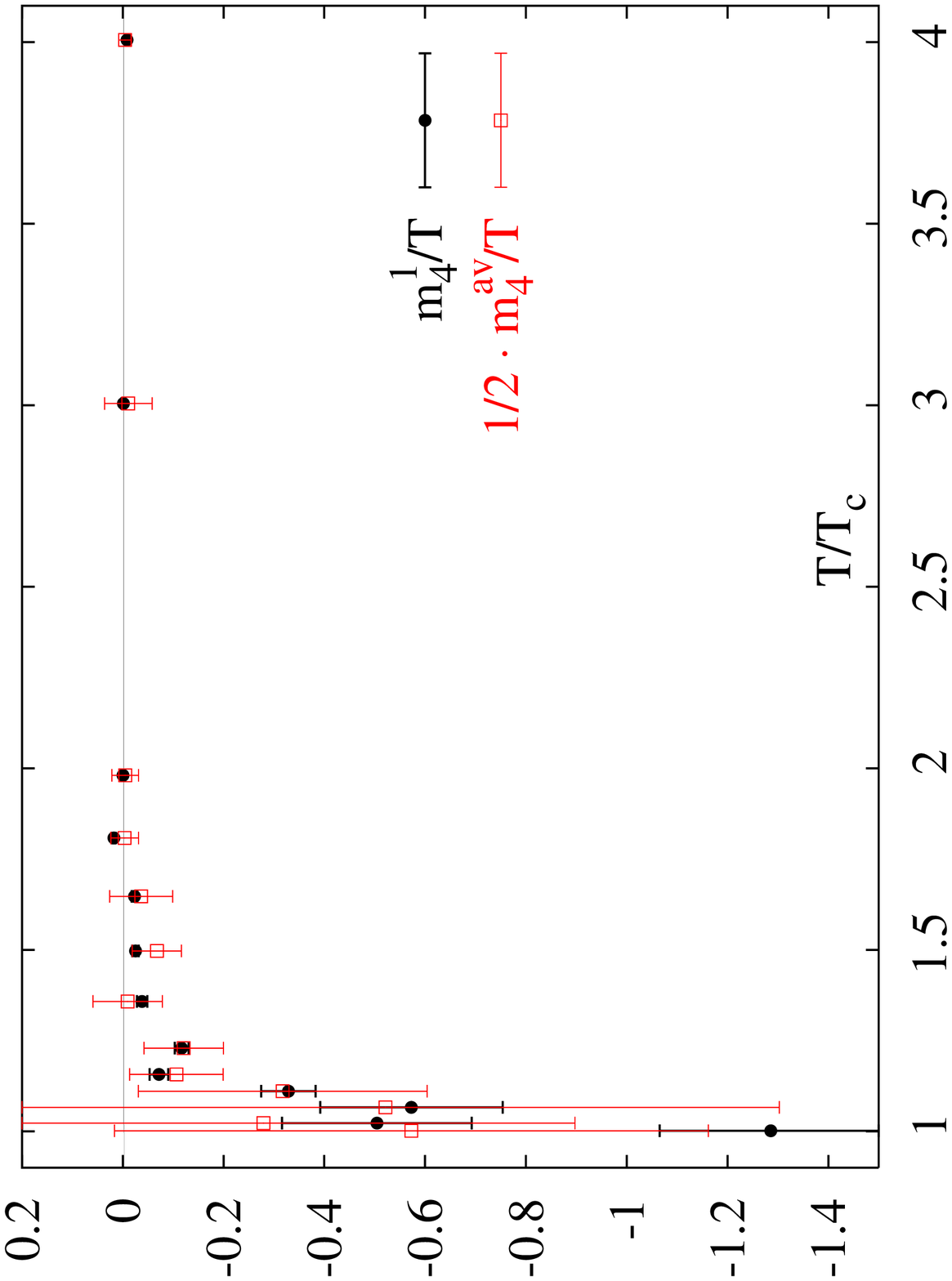} \end{turn} &
\begin{turn}{270} \epsfbox{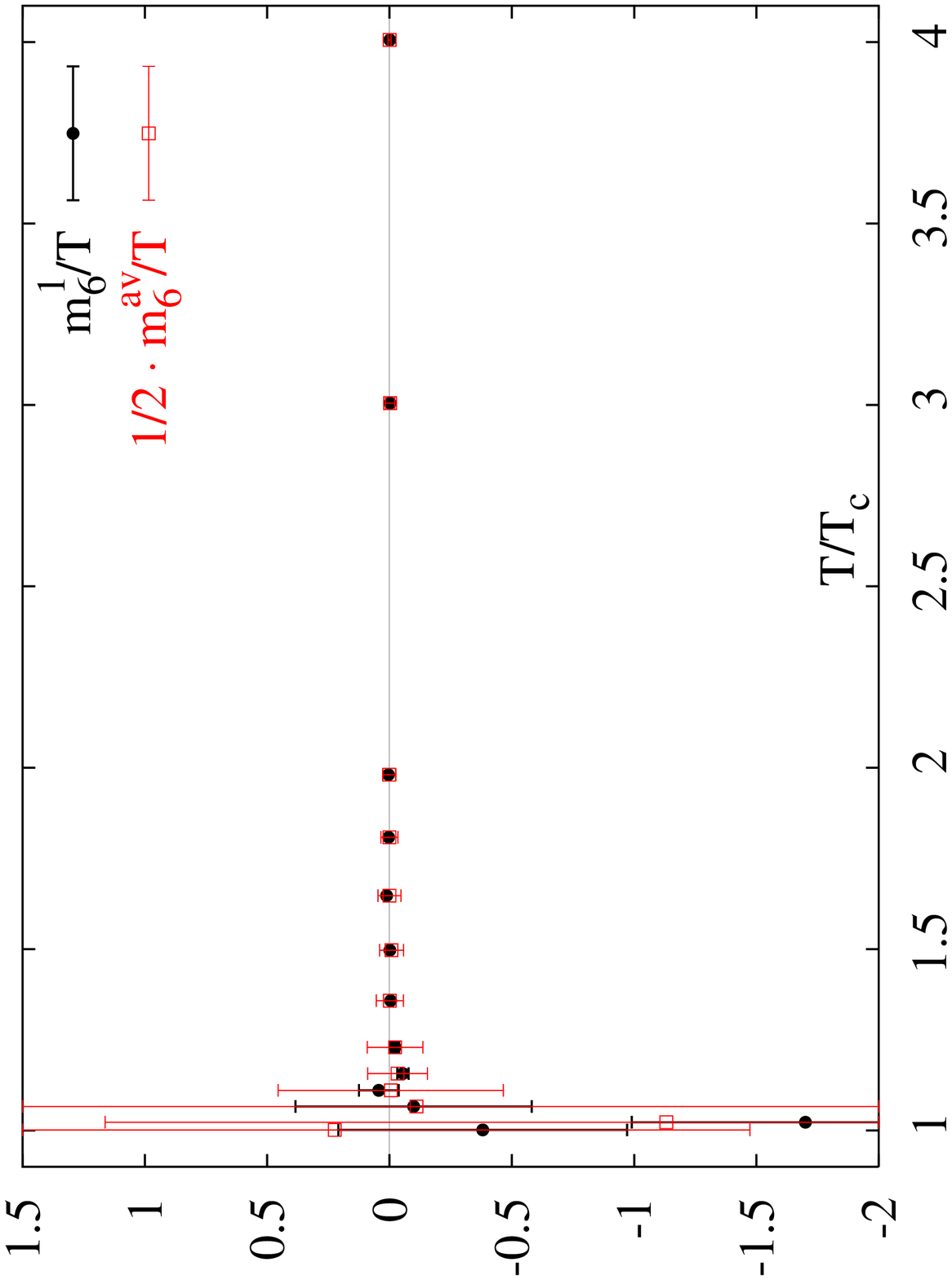} \end{turn}\\
\mbox{(c)} & \mbox{(d)}
\end{array}
\eew
\caption{Expansion coefficients of screening masses in the colour singlet
($m^1_n$) and colour averaged ($m^{\rm av}_n$) channels. Except for $n=0$
$m^x_n(T)$ is determined from the $r \to \infty$ limit of (\ref{debye_4}). The
lines are the first order perturbative predictions according to
eq.~\ref{m_lim}. The dotted lines in (a) show the 1$\sigma$-range of the $\chi^2$-fit with parameter $A$.\label{mass2}}
\end{figure}
\begin{figure}[t]
\bew
\begin{array}{cc}
\begin{turn}{270} \epsfbox{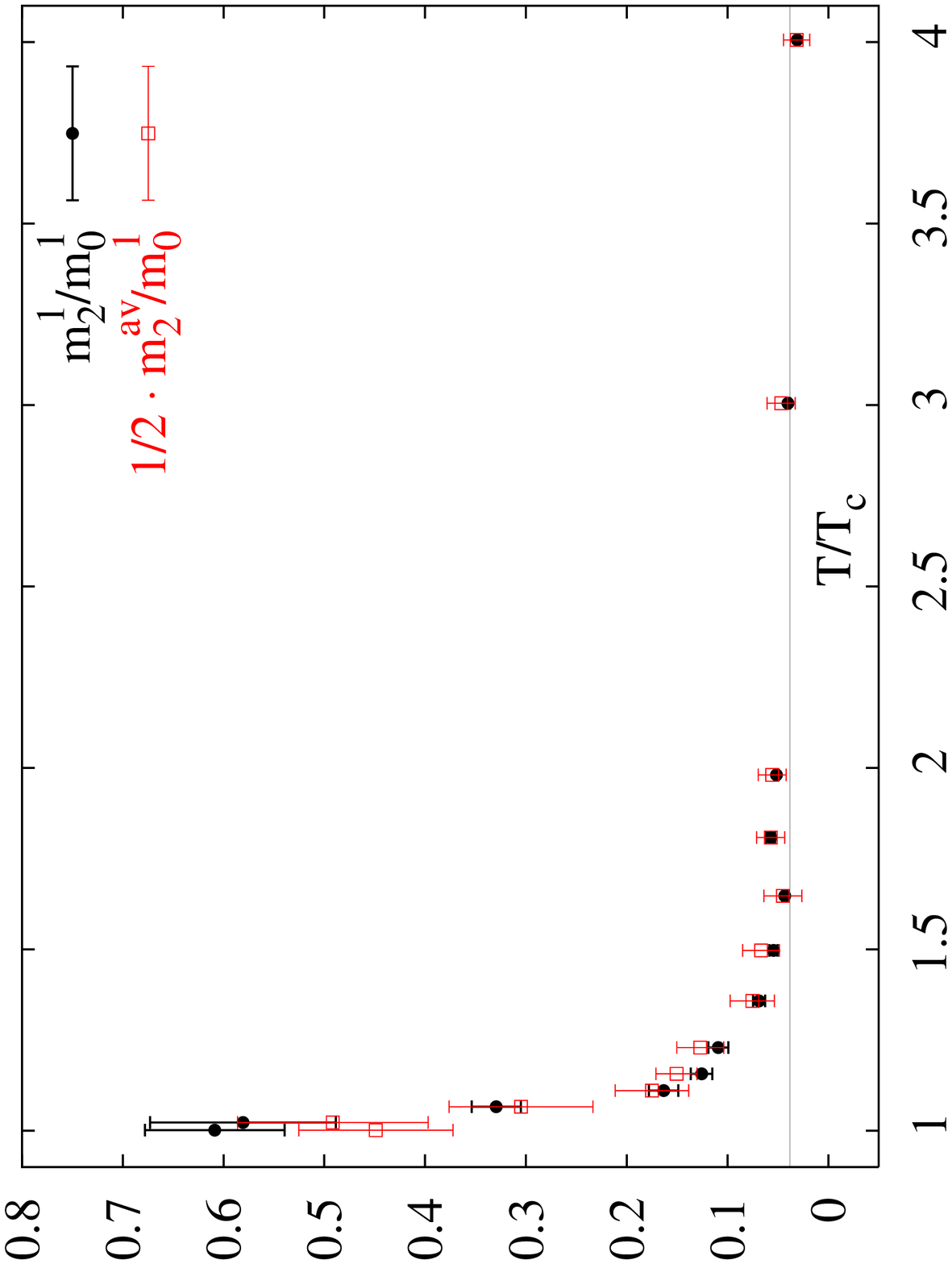} \end{turn} &
\begin{turn}{270} \epsfbox{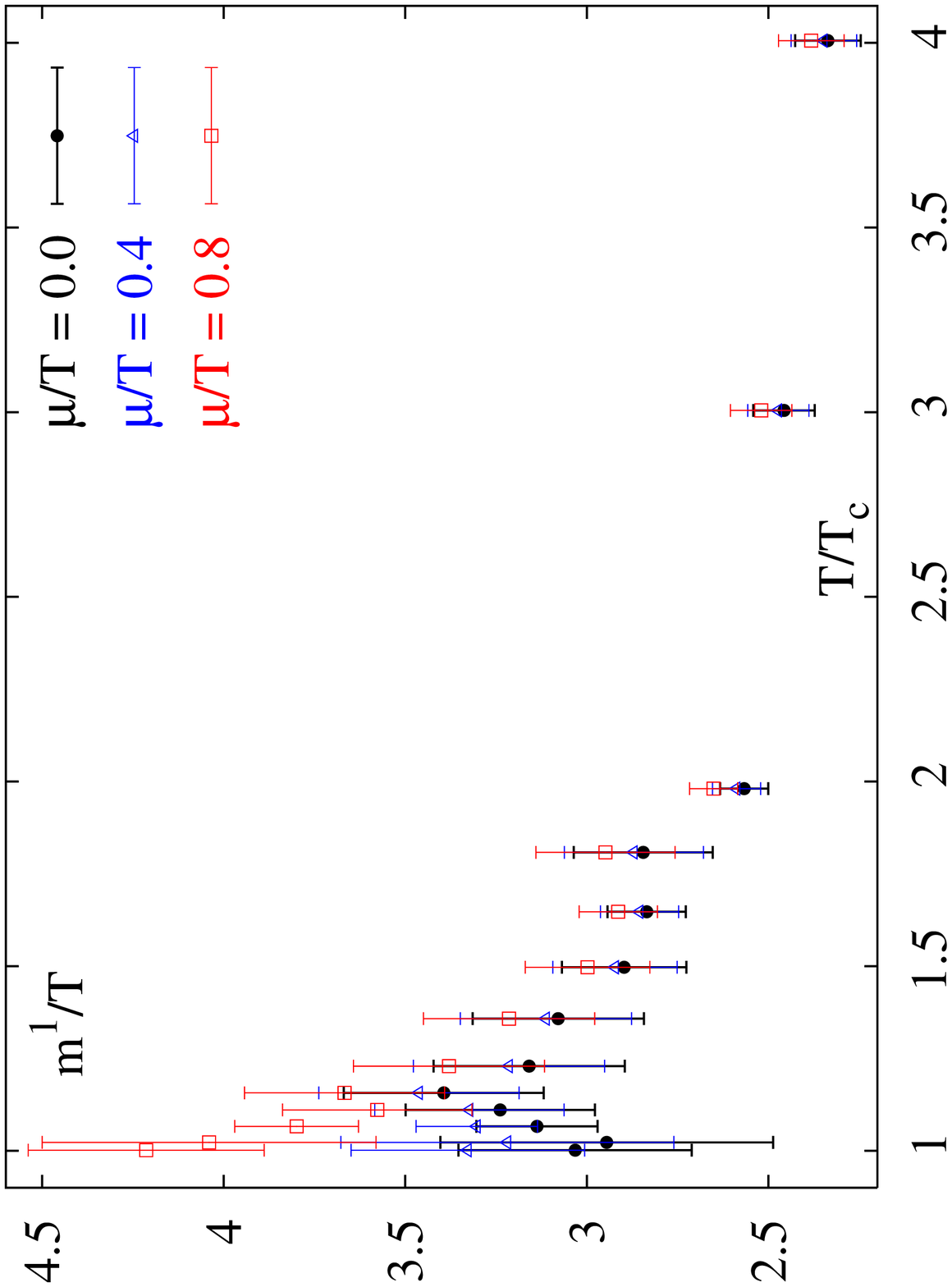} \end{turn}\\
\mbox{(a)} & \mbox{(b)}
\end{array}\\
\eew
\caption{The ratio $m^x_2(T)/m^1_0(T)$ (a) and the screening mass $m^1(T)$
  evaluated for $\mu/T = 0.0, 0.4$ and $0.8$ including the $0^{th}$ and $2^{nd}$ order
  expansion coefficients in $\mu$ (b). The line in (a) shows the leading order
  perturbative prediction, which is $3/8\pi^2$.
\label{m2m0_2}}
\end{figure}

\begin{eqnarray}
\Delta F^{\rm av,1}_{Q\bar{Q}}(r, T, \mu) &=& 
F^{\rm av,1}_{Q\bar{Q}}(\infty, T, \mu) - F^{\rm av,1}_{Q\bar{Q}}(r, T,\mu) 
\; , \nonumber\\
&\sim& \f{1}{r^n}e^{-m^{\rm av,1}(T,\mu) r} 
\label{delta}
\end{eqnarray}
with $n=1, 2$ for the singlet and colour averaged free energies respectively.
In the infinite distance limit we thus can extract the screening masses, 
\be
m^{\rm av,1}(T,\mu) = -\lim_{r\rightarrow \infty} \frac{1}{r} \ln \left( 
\Delta F^{\rm av,1}_{Q\bar{Q}}(r, T,\mu) \right)
\; . \label{debye-ansatz}
\ee
We use this as our starting point to derive a Taylor expansion for 
the screening masses. Expanding the logarithm in eq.~\ref{debye-ansatz} 
in powers of $\mu/T$ it is obvious that also the screening masses are
even functions in $\mu/T$, 
\be
m^{\rm av,1}(T, \mu) &=& m^{\rm av,1}_0(T) + m^{\rm av,1}_2(T) 
\left(\f{\mu}{T}\right)^2 +
m^{\rm av,1}_4(T)\left(\f{\mu}{T}\right)^4 + m^{\rm av,1}_6(T) 
\left(\f{\mu}{T} \right)^6
+ {\cal O}\left(\left(\f{\mu}{T}\right)^8\right) \; .
\label{ansatz}
\ee
To analyze the approach of the various expansion coefficients to the large 
distance limits we introduce effective masses, 
$m_{{\rm eff},n}^x (r, T)$, with $x= {\rm av}, 1$,
\bs
m_{{\rm eff},2}^x(r, T) &=& - \f{1}{r} 
\frac{\Delta f^x_{Q\bar{Q},2}(r, T)}{\Delta f^x_{Q\bar{Q},0}(r, T)} \;, \\
m_{{\rm eff},4}^x(r, T) &=& - \f{1}{r} \left[\f{\Delta 
f^x_{Q\bar{Q},4}(r, T)}{\Delta f^x_{Q\bar{Q},0}(r, T)} - 
\f{1}{2} \left(\f{\Delta f^x_{Q\bar{Q},2}(r, T)}{\Delta 
f^x_{Q\bar{Q},0}(r, T)}\right)^2\right]\; ,\\
m_{{\rm eff},6}^x(r, T) &=& - \f{1}{r} \left[\f{\Delta 
f^x_{Q\bar{Q},6}(r, T)}{\Delta f^x_{Q\bar{Q},0}(r, T)} -
  \f{\Delta f^x_{Q\bar{Q},4}(r, T)\Delta 
f^x_{Q\bar{Q},2}(r, T)}{\Delta f^x_{Q\bar{Q},0}(r, T)^2} + 
\f{1}{3} \left(\f{\Delta f^x_{Q\bar{Q},2}(r, T)}{\Delta 
f^x_{Q\bar{Q},0}(r, T)}\right)^3\right] \; .
\label{debye_4}
\es
In the limit of large distances these relations define the expansion
coefficients of the colour averaged and singlet screening masses,
\be
m^{\rm av,1}_n(T)=\lim_{r\rightarrow \infty} 
m_{{\rm eff},n}^{\rm av,1}(r, T) \; .\label{limav}
\ee
As will become obvious in the following the effective masses defined
above show only little $r$-dependence. They are thus suitable for
a determination of the $\mu$-dependent corrections to the screening
masses. This is not the case for the leading order, $\mu$-independent,
contribution.
In order to determine $m^1_0(T)$ we use an ansatz for the large distance
behaviour of the singlet free energy motivated by leading order high
temperature perturbation theory,
\be
f^1_{Q\bar{Q}, 0}(r, T) = f^1_{Q\bar{Q},0}(\infty, T) - 
\f{4}{3}\f{\alpha_0(T)}{r} e^{-m^1_0(T) r}.
\ee
We fit our data to this equation using $\alpha_0(T)$ and $m^1_0(T)$ as fit
parameters where $f^1_{Q\bar{Q},0}(\infty,T)$ is determined as described in the
previous section. We choose the same fitting procedure as 
in \cite{fthq2} namely averaging results received from five fit windows with 
left borders between $rT = 0.8$ and $rT = 1.0$ and right border at $rT=1.73$.
While the above ansatz is known to describe rather well the large
distance behaviour of the color singlet free energy, it also is known
that the sub-leading power-like corrections are much more difficult to
control in the case of the colour averaged free energy. For this reason
we will analyze here only the leading order contribution to the singlet
screening mass.

Results for effective masses in the singlet channel 
are shown in Fig.~\ref{df2df0} as function of $rT$ for one value of the 
temperature. As can be seen the asymptotic value is indeed reached quickly 
before the errors grow large at distances $rT\gsim 1$. 
The expansion coefficients
$m^{\rm av,1}_2(T)$, $m^{\rm av,1}_4(T)$ and $m^{\rm av,1}_6(T)$ are thus 
well determined from the plateau
values of these ratios. Similar results hold in the colour averaged channel. We
found the left border of the plateau to lie between $rT = 0.48$ close to $T_c$
and $rT = 0.23$ for $T>1.15T_c$. Results for the various expansion
coefficients are shown in Fig.~\ref{mass2}. This
figure shows that at high temperatures the $\mu$-dependent corrections
to the screening mass of the colour averaged
free energies $m^{\rm av}(T,\mu)$ are twice as large as those of the (Debye) 
screening
mass in the singlet channel, $m^1(T,\mu)$. This is expected from perturbation 
theory, which suggests that the leading order contribution to the colour 
singlet free energy is given by one gluon exchange while the colour averaged 
free energy is dominated by two gluon exchange. 
Using resummed gluon propagators then 
leads to screening masses that differ by a factor of 2,
\be
m^1_n(T) = \f{1}{2} m^{\rm av}_n(T)\;, \quad n = 2, 4, 6 \; ,
\label{half}
\ee
Our results suggest that this relation holds already close to $T_c$
(Fig.~\ref{mass2}).
We thus have no evidence for
large contributions from the magnetic sector, which is expected to
dominate the screening in the colour averaged channel at asymptotically
large temperatures \cite{Yaffe} and which would violate the simple 
relation given in eq.~\ref{half}.

In order to compare the expansion coefficients with perturbation theory we need
to specify the running coupling $g(T)$. Following \cite{fthq2} we use the
next-to-leading order perturbative result for the running of the coupling with
temperature but allow for a free overall scale factor. We thus fit our data 
on the $T$-dependence of the leading order ($\mu=0$) screening mass by the ansatz,
\be
m^1_0(T) = A \cdot \f{2}{\sqrt{3}} g(T)T \; ,\label{am0}
\ee
with the 2nd order perturbative running coupling,
\be
g(T)^2 = \left[ \f{29}{24\pi^2} \ln{\left(\f{\widetilde{\mu} T}{\Lambda_{\overline{MS}}}\right)} +
  \f{115}{232 \pi^2} \ln{\left(\ln{\left(\f{\widetilde{\mu} T}{\Lambda_{\overline{MS}}}\right)}\right)}\right]^{-1}\;,
\ee
where we use $T_c/\Lambda_{\overline{MS}} = 0.77(21)$ and the scale
$\widetilde{\mu} = 2\pi$ as in ~\cite{fthq2}. Fitting our data to eq.~\ref{am0} with fit parameter $A$, yields
\bs
A &=&  1.397(18)\;,
\es
which is almost identical to the result in \cite{fthq2} where the data for
$T=3T_c$ were still missing. Our fit result is included in Fig.~\ref{mass2}. We
also compare the temperature dependence of 
$m^1_2(T)$, $m^1_4(T)$ and $m^1_6(T)$ 
with corresponding expansion coefficients of the perturbative Debye mass
which result from an expansion of 
(\ref{pert}) using (\ref{am0}) as the 0th order. 
These expansion coefficients are alternating in sign,
\bs
m_{D,2}(T) &=& \f{\sqrt{3}}{4\pi^2} \cdot A g(T)\;,\\
m_{D,4}(T) &=& -\f{3\sqrt{3}}{64\pi^4} \cdot A g(T)\;,\\
m_{D,6}(T) &=& \f{9\sqrt{3}}{512\pi^6} \cdot A g(T)\;. \label{m_lim}
\es
At least for the second order coefficient $m^1_{2}(T)$ we find that this 
yields a satisfactory description of the numerical results for 
$T \gsim 2 T_c$. Eq.~\ref{m_lim}
shows that subsequent terms differ by about an order of magnitude, which
explains why our signal for a non-zero contribution $m^1_n(T)$ is rather 
poor for $n>2$.

From (\ref{m_lim}) we find $m_{D,2}(T)/m_{D,0}(T) = 3/8\pi^2$ which is 
independent of $A$ and $g(T)$ and is compared with our numerical results in
Fig.~\ref{m2m0_2}(a). We note that the perturbative value for this ratio is
already reached for $T/T_c \gsim 2$. 
In Fig.~\ref{m2m0_2}(b) we show the $\mu$-dependence of the singlet
screening mass for a small 
values of $\mu/T$. Here we included only contributions from the 0th and 2nd 
order expansion in the calculation of $m^1(\mu,T)/T$.

\section{Conclusions}
\label{conclusions}
We have analyzed the response of colour singlet and colour averaged
heavy quark free energies to a non-vanishing
baryon chemical potential and have calculated the resulting dependence of 
screening masses  on the chemical potential. Using a Taylor expansion in
$\mu/T$ we get stable results for the leading, non-vanishing correction,
$m^1_2(T)$, which is ${\cal O}((\mu/T)^2)$. We find that this correction in
absolute units as well as its ratio with the leading order screening mass,
$m^1_0(T)$, is large in the vicinity of the transition temperature. The ratio
$m^1_2(T)/m^1_0(T)$ is in agreement with perturbation theory for $T\gsim 2T_c$
indicating that the expansion coefficients $m^1_n(T)$ 
receive the same multiplicative rescaling as the leading 
order screening mass.

A calculation of the $\mu$-dependent corrections to the
screening mass in the colour averaged channel shows that these corrections
are twice as large as those in the color singlet channel for all 
temperatures $T > T_c$. This agreement with leading order perturbation 
theory indeed is quite remarkable as it suggests that the leading 
contribution to the $\mu$-dependent corrections of the colour averaged
screening mass 
is due to 
two-gluon exchange.


The higher order expansion coefficients of the screening mass vanish within
statistical errors at temperatures larger than $1.2 T_c$. The analysis of the
asymptotic behaviour of the free energies themselves, however, suggests that
these corrections are non-zero but small at high temperature and have
alternating signs. This is consistent with the leading order perturbative result
for the Debye mass subsequent expansion coefficients of which drop
by more than an order of magnitude and alternate in sign as they arise from an
expansion of a square root.

Our results thus suggest that at least for small values of the chemical 
potential and fixed temperature the screening length in a baryon rich 
quark gluon plasma decreases with increasing value of the chemical potential. 
This is consistent with the expectation that the transition to the high 
temperature phase shifts to lower temperatures
at non-zero baryon chemical potential.

\section{Acknowledgments}
\label{ackn}
This work has been supported partly through the DFG under grant KA 1198/6-4, the
GSI collaboration grant BI-KAR and a grant of the BMBF under contract no. 06BI106. MD is supported through a fellowship of the DFG funded graduate school GRK 881. 
The work of FK has been partly supported by a contract DE-AC02-98CH1-886 with
the U.S. Department of Energy.

\begin{appendix}
\section{Calculation of expansion coefficients}
\label{calc}
The $\mu$-dependent expectation value of a complex quantity ${\cal O}$ is
\be
\left< {\cal O} \right>_\mu = \f{1}{Z_\mu} \int DU {\cal O} \Delta e^{-S} =
\f{\int DU {\cal O} \Delta e^{-S}}{\int DU \Delta e^{-S}}\;, \label{expec}
\ee
where $Z_\mu$ is the partition function for finite $\mu$ and where $\Delta =
\left(\det{M}(\mu)\right)^{N_f/4}$ is the determinant of the fermion matrix. In
the following we denote expectation values for vanishing $\mu$ as $\left< \cdots\right> = \left<\cdots\right>_0$. We define
\be
L_n \equiv \left. \td{^n\ln{\Delta}}{\mu^n} \right|_{\mu = 0} = \left.\f{N_f}{4} \td{^n\ln{\det{M}}}{\mu^n} \right|_{\mu = 0}.
\ee
The $L_n$ can be written as traces over the inverse of the fermion matrix and 
its derivatives
\be
L_1 &=& \left. \f{N_f}{4} \mbox{Tr}\left(\pdm\right) \right|_{\mu=0},\\
L_2 &=& \left. \f{N_f}{4} \mbox{Tr}\left(\pdmm{2} - \pdm\pdm\right) \right|_{\mu=0},\\
L_3 &=& \left. \f{N_f}{4} \mbox{Tr}\left(\pdmm{3} - 3\pdm\pdmm{2}
    + 2\pdm\pdm\pdm\right) \right|_{\mu=0},\\
L_4 &=& \f{N_f}{4} \mbox{Tr}\left(\pdmm{4} - 4\pdm\pdmm{3} -
  3\pdmm{2}\pdmm{2} + 12\pdm\pdm\pdmm{2}\right.\nnn
&& \quad \left.\left. - 6\pdm\pdm\pdm \pdm\right)\right|_{\mu = 0},\\
L_5 &=& \f{N_f}{4} \mbox{Tr}\left(\pdmm{5} - 5\pdm\pdmm{4} -
  10\pdmm{2}\pdmm{3} + 20\pdm\pdm\pdmm{3}\right.\nnn
&& \quad + 30\pdm\pdmm{2}\pdmm{2} - 60 \pdm\pdm\pdm \pdmm{2}\nnn
&& \quad \left.\left. + 24\pdm\pdm \pdm\pdm\pdm\right)\right|_{\mu=0},\\
L_6 &=& \f{N_f}{4} \mbox{Tr}\left(\pdmm{6} - 6\pdm\pdmm{5} - 15\pdmm{2}\pdmm{4} - 10\pdmm{3}\pdmm{3} \right.\nnn
&& \quad+30\pdm\pdm\pdmm{4} + 60\pdm \pdmm{2}\pdmm{3} +  60\pdmm{2}\pdm\nnn
&& \quad \pdmm{3} + 30\pdmm{2}\pdmm{2}\pdmm{2} - 120\pdm\pdm\pdm\pdmm{3} \nnn
&& \quad - 180\pdm\pdm\pdmm{2}\pdmm{2} - 90\pdm\pdmm{2}\pdm\pdmm{2} \nnn
&& \quad + 360\pdm\pdm\pdm\pdm \pdmm{2} - 120\pdm\pdm\pdm\nnn
&& \quad \left.\left.\pdm\pdm\pdm\right)\right|_{\mu=0}.
\ee
From
\be
M^\dagger(\mu) &=& \gamma_5 M(-\mu) \gamma_5
\ee
it follows that $L_n$ is real for even and imaginary for odd $n$. Using $\Delta = e^{\ln{\Delta}}$ we find
\be
\Delta(\mu) &=& \Delta(0) \left( 1 + D_1 \mu + D_2 \mu^2 + \cdots + D_6 \mu^6 
+ {\cal O}(\mu^7) \right)\;,
\ee
where
\bs
D_1 &=& L_1\;,\\
D_2 &=& \f{1}{2} \left(L_1^2 + L_2 \right)\;,\\
D_3 &=& \f{1}{6} \left( L_1^3 + 3 L_1 L_2 + L_3 \right)\;,\\
D_4 &=& \f{1}{24} \left( L_1^4 + 6 L_1^2 L_2 + 3 L_2^2 + 4 L_1 L_3 + L_4
  \right),\\
D_5 &=& \f{1}{120} \left( L_1^5 + 10 L_1^3 L_2 + 15 L_1 L_2^2 + 10 L_1^2 L_3
+ 10 L_2 L_3 + 5 L_1 L_4 + L_5 \right)\;,\\
D_6 &=& \f{1}{720} \left( L_1^6 + 15 L_1^4 L_2 + 45 L_1^2 L_2^2 + 15 L_2^3 +
    20 L_1^3 L_3 + 60 L_1 L_2 L_3\right.\nnn
&& \quad \left. + 10 L_3^2 + 15 L_1^2 L_4 + 15 L_2 L_4 + 6 L_1 L_5 + L_6 \right)\;. 
\es
We immediately see that $D_n$ is real for even and imaginary for odd $n$. Because
\be
Z_\mu = \left< 1 + D_1 \mu + \cdots D_6 \mu^6 \right> + {\cal O}(\mu^7)
\ee
is real one has $\left< D_n \right> = 0$ for odd $n$. We consider the case 
where the
observable ${\cal O}$ is independent of $\mu$. The expectation value 
(\ref{expec}) then becomes
\be
\left< {\cal O} \right>_\mu &=& \f{\left< {\cal O} \right> + \left< {\cal O} D_1
  \right> \mu + \cdots + \left< {\cal O} D_6 \right> \mu^6}{1 + \left< D_2
  \right> \mu^2 + \cdots + \left< D_6 \right> \mu^6} + {\cal O}(\mu^7)\;.
\ee
Expanding in powers of $\mu$ we get
\be
\left< {\cal O} \right>_\mu &=& \left< {\cal O} \right> \left( 1 + {\cal O}_1
  \mu + \left( - {\cal D}_2 + {\cal O}_2 \right) \mu^2 + \left( - {\cal D}_2 {\cal O}_1 + {\cal O}_3 \right) \mu^3 + \left( {\cal D}_2^2 - {\cal D}_4 - {\cal
    D}_2 {\cal O}_2 + {\cal O}_4 \right) \mu^4 + \left( {\cal D}_2^2 {\cal O}_1
  - {\cal D}_4 {\cal O}_1\right.\right. \nnn
&& \quad \left. - {\cal D}_2 {\cal O}_3 + {\cal O}_5
  \right) \mu^5 + \left( - {\cal D}_2^3 + 2 {\cal D}_2 {\cal D}_4 - \left. {\cal D}_6 + {\cal
      D}_2^2 {\cal O}_2 - {\cal D}_4 {\cal O}_2 - {\cal D}_2 {\cal O}_4 + {\cal
      O}_6 \right) \mu^6 \right) + {\cal O}(\mu^7) \label{expan}
\ee
where we use the notation
\bs
{\cal O}_i &=& \f{\left< {\cal O} D_i \right>}{\left< {\cal O} \right>}\;,\\
{\cal D}_i &=& \left< D_i \right>\;.
\es
In the case that ${\cal O}$ is strictly real on every configuration 
${\cal O}D_n$
is imaginary for odd n. In order to keep $\left<{\cal O}\right>_\mu$ real
$\left<{\cal O} D_n\right>$ has to vanish for odd $n$ and the preceding 
expansion simplifies to
\be
\left< {\cal O} \right>_\mu &=& \left< {\cal O} \right> \left( 1 + \left( -
    {\cal D}_2 + {\cal O}_2 \right) \mu^2 + \left( {\cal D}_2^2 - {\cal D}_4 -
    {\cal D}_2 {\cal O}_2 + {\cal O}_4 \right) \mu^4\right.\nnn
&& + \left.\left( - {\cal D}_2^3 + 2 {\cal D}_2
  {\cal D}_4 - {\cal D}_6 + {\cal D}_2^2 {\cal O}_2 - {\cal D}_4 {\cal O}_2 - {\cal D}_2 {\cal O}_4 + 
{\cal O}_6 \right) \mu^6 \right) + {\cal O}(\mu^8)\;,
\ee
i.e. this formula is applicable to the correlation function in eq.~\ref{corrs}. 

Because free energies are calculated from logarithms of correlation functions we
give here the expansion coefficients of the logarithm of an observable ${\cal
  O}$ which can be obtained by inserting the above expansion into the expansion
of the logarithm. For a generic, not necessarily real observable the 
expansion is
\be
\ln{\left< {\cal O} \right>_\mu} &=& \ln{\left< {\cal O} \right>} + {\cal O}_1 \mu + \left( -
{\cal D}_2 - \f{1}{2} {\cal O}_1^2 + {\cal O}_2 \right) \mu^2 + \left( \f{1}{3} {\cal O}_1^3 - {\cal O}_1 {\cal O}_2 + {\cal O}_3 \right)
\mu^3 +\left(\f{1}{2} {\cal D}_2^2 - {\cal D}_4 -\right.\nnn
&& \quad \left. \f{1}{4} {\cal O}_1^4 + {\cal O}_1^2 {\cal O}_2 - \f{1}{2} {\cal O}_2^2 - {\cal O}_1 {\cal O}_3 + {\cal O}_4
\right) \mu^4 + \left( \f{1}{5} {\cal O}_1^5 - {\cal O}_1^3 {\cal O}_2 + {\cal O}_1 {\cal
    O}_2^2 + {\cal O}_1^2 {\cal O}_3 - {\cal O}_2 {\cal O}_3\right.\nnn
&& \quad \left. - {\cal O}_1 {\cal O}_4 + {\cal O}_5 \right) \mu^5 + \left( - \f{1}{3} {\cal D}_2^3 + {\cal D}_2 {\cal D}_4
  - {\cal D}_6 - \f{1}{6} {\cal O}_1^6 + {\cal O}_1^4 {\cal O}_2 - \f{3}{2}
  {\cal O}_1^2 {\cal O}_2^2 + \f{1}{3} {\cal O}_2^3 - {\cal O}_1^3 {\cal O}_3 \right.\nnn
&& \quad \left. + 2 {\cal O}_1 {\cal O}_2 {\cal O}_3 - \f{1}{2} {\cal O}_3^2 + {\cal
    O}_1^2 {\cal O}_4 - {\cal O}_2 {\cal O}_4 - {\cal O}_1 {\cal O}_5 + {\cal
    O}_6 \right) \mu^6 + {\cal O}(\mu^7)\;.
\ee

\noindent
For real observables this reduces to
\be
\ln{\left< {\cal O} \right>_\mu} &=& \ln{\left< O \right>} + \left( - {\cal D}_2 +
  {\cal O}_2 \right) \mu^2 +\left(\f{1}{2} {\cal D}_2^2 - {\cal D}_4
  - \f{1}{2} {\cal O}_2^2 + {\cal O}_4 \right) \mu^4\nnn
&& \quad + \left( - \f{1}{3} {\cal D}_2^3 + {\cal D}_2 {\cal D}_4 - {\cal D}_6 + \f{1}{3} {\cal O}_2^3 - {\cal O}_2 {\cal O}_4 + {\cal O}_6 \right) \mu^6 +
{\cal O}(\mu^8)\;. \label{realobs}
\ee
\end{appendix}


\begin{thebibliography}{99}

\bibitem{review}
F.~Karsch and E.~Laermann, hep-lat/0305025.
\bibitem{fthq}
O.~Kaczmarek, F.~Karsch, F.~Zantow, P.~Petrecky, Phys.Rev. {\bf D 70} (2004) 074505.
\bibitem{fthq2}
O.~Kaczmarek, F.~Zantow, Phys. Rev. D71 (2005) 114510.
\bibitem{pisarski}
A.~Dumitru, Y.~Hatta, J.~Lenaghan, K.~Orginos and R.~D.~Pisarski,
Phys.\ Rev.\ D {\bf 70} (2004) 034511.
\bibitem{vuorinen} 
A.~Vuorinen,
Phys.\ Rev.\ D {\bf 68} (2003) 054017.
\bibitem{rebhan}
A.~Rebhan and P.~Romatschke, Phys.\ Rev.\ D {\bf 68} (2003) 025022.
\bibitem{debye-mu}
T.~Toimela,
Phys.\ Lett.\ B {\bf 124} (1983) 407. 
\bibitem{fodor-debye}
Z.~Fodor, S.~D.~Katz, K.~K.~Szabo and A.~I.~T\'oth,
Nucl.\ Phys.\ B (Proc.\ Suppl.)\ {\bf 140} (2005) 508.
\bibitem{eos}
C.~R.~Allton, S.~Ejiri, S.~J.~Hands, O.~Kaczmarek, F.~Karsch,
E.~Laermann and C.~Schmidt,
Phys.\ Rev.\ D {\bf 68} (2003) 014507;\\
C.~R.~Allton, M.~Doring, S.~Ejiri, S.J.~Hands, O.~Kaczmarek,
F.~Karsch, E.~Laermann, K.~Redlich, Phys. Rev D 71 (2005) 054508.
\bibitem{Fodor} 
Z.~Fodor, S.~D.~Katz and K.~K.~Szabo,
Phys.\ Lett.\ B {\bf 568} (2003) 73;\\
F.~Csikor, G.~I.~Egri, Z.~Fodor, S.~D.~Katz, K.~K.~Szabo and A.~I.~T\'oth,
JHEP {\bf 0405} (2004) 046.
\bibitem{lombardo}
M.~D'Elia and M.~P.~Lombardo,
Phys.\ Rev.\ D {\bf 70} (2004) 074509.
\bibitem{allton}
C.~R. Allton, S.~Ejiri, S.~J.~Hands, O.~Kaczmarek, F.~Karsch, E.~Laermann, Ch.~Schmidt and L.~Scorzato, Phys. Rev. D 66 (2002) 074507.
\bibitem{kacz}
O.~Kaczmarek, F.~Karsch, P.~Petrecky, F.~Zantow, Phys. Lett. {\bf B 543} (2002) 41.
\bibitem{peikert}
F.~Karsch, E.~Laermann, A.~Peikert, Nucl.\ Phys.\ B {\bf 605} (2001) 579.
\bibitem{landau} L.D. Landau, E.M. Lifshitz, {\it Statistical Physics}; E. Gava, R. Jengo, Phys. Lett. {\bf B 105} (1981) 285
\bibitem{Yaffe}
P. Arnold and L.G. Yaffe, Phys. Rev. D {\bf 52} 7208 (1995).
\end{thebibliography}
\end{document}